\documentclass[draft]{agujournal2019}
\usepackage{float}
\usepackage{url}
\usepackage{lineno}
\usepackage[inline]{trackchanges}
\usepackage{soul}
\usepackage{amsmath}
\usepackage{amssymb}
\usepackage{ragged2e}
\usepackage{algorithm}
\usepackage{algpseudocode}
\usepackage{comment}
\usepackage{natbib}

\draftfalse

\begin{document}
\justifying

\title{Bayesian Full-waveform Monitoring of CO$_2$ Storage with Fluid-flow Priors via Generative Modeling}
\authors{Haipeng Li\affil{1},  Nanzhe Wang\affil{2,3}, Louis J. Durlofsky\affil{2}, Biondo L. Biondi\affil{1}}
\affiliation{1}{Department of Geophysics, Stanford University, CA 94305, USA.}
\affiliation{2}{Department of Energy Science and Engineering, Stanford University, CA 94305, USA.}
\affiliation{3}{Now at Institute of GeoEnergy Engineering, School of Energy, Geoscience, Infrastructure and Society, Heriot-Watt University, Edinburgh, EH14 4AS, UK.}

\correspondingauthor{Nanzhe Wang}{nanzhe.wang@hw.ac.uk}

\begin{keypoints}
\item A Bayesian seismic full-waveform monitoring framework integrates geostatistics, fluid-flow simulations, and rock physics for CO$_2$ storage.
\item A variational autoencoder learns a compact generative prior and enables efficient Hamiltonian Monte Carlo sampling in latent space.
\item Posterior ensembles yield robust plume monitoring under sparse, noisy data and quantify uncertainty for survey design and bias correction.
\end{keypoints}

\begin{abstract}
% what is the content
% what we already have and the problems.
% what we want and how to solve it
% Findings
% Conclusions
% Perspectives
Quantitative monitoring of subsurface changes is essential for ensuring the safety of geological CO$_2$ sequestration. Full-waveform monitoring (FWM) can resolve these changes at high spatial resolution, but conventional deterministic inversion lacks uncertainty quantification and incorporates only limited prior information. Deterministic approaches can also yield unreliable results with sparse and noisy seismic data. To address these limitations, we develop a Bayesian FWM framework that combines reservoir flow physics with generative prior modeling. Prior CO$_2$ saturation realizations are constructed by performing multiphase flow simulations on prior geological realizations. Seismic velocity is related to saturation through rock physics modeling. A variational autoencoder (VAE) trained on the priors maps high-dimensional CO$_2$ saturation fields onto a low-dimensional, approximately Gaussian latent space, enabling efficient Bayesian inference while retaining the key geometrical structure of the CO$_2$ plume. Hamiltonian Monte Carlo (HMC) is used to infer CO$_2$ saturation changes from time-lapse seismic data and to quantify associated uncertainties. Numerical results show that this approach improves inversion stability and accuracy under extremely sparse and noisy acquisition, whereas deterministic methods become unreliable. Statistical seismic monitoring provides posterior uncertainty estimates that identify where additional measurements would most reduce ambiguity and mitigate errors arising from biased rock physics parameters. The framework combines reservoir physics, generative priors, and Bayesian inference to provide uncertainty quantification for time-lapse monitoring of CO$_2$ storage and other subsurface processes.
\end{abstract}

\section*{Plain Language Summary}
Geological carbon storage is a critical component for reducing greenhouse gas emissions by injecting CO$_2$ deep into the subsurface. To ensure safe storage, it is essential to monitor how the injected CO$_2$ evolves. Seismic monitoring provides detailed images of this process, but traditional approaches yield only a single best-fit model and do not indicate how uncertain that image may be, which is important for real-world decisions. We develop a Bayesian monitoring framework integrating geostatistical data, reservoir simulations, and generative modeling. We first generate an ensemble of geomodels to represent geological variations and simulate CO$_2$ migration with fluid flow physics. These plumes are learned with a neural network, which captures plausible CO$_2$ plume patterns in a compact form. Bayesian inference then combines this prior information with time-lapse seismic measurements, allowing estimation of both the most likely CO$_2$ distribution and its uncertainty. Our results show that this probabilistic approach produces more accurate and stable performance, even with noisy data and only a single seismic source. It also helps identify where additional measurements would most reduce uncertainty. Overall, this framework integrates multi-source data within a Bayesian workflow and offers a reliable probabilistic approach for monitoring CO$_2$ storage and other subsurface processes.
\newpage

\section{Introduction}
% Background and Motivation for Geological Carbon Sequestration
Geological carbon sequestration (GCS) mitigates anthropogenic CO$_2$ emissions by securely storing the CO$_2$ in deep geological formations~\citep{davis2019geophysics}. Ensuring the long-term safety and efficiency of this operation requires reliable subsurface monitoring to verify that the injected CO$_2$ remains trapped, migrates as predicted, and does not leak into unintended formations~\citep{jenkins2020state}. Quantitative and high-resolution monitoring of injected CO$_2$ is therefore essential for verifying storage integrity, constraining fluid-flow simulations, and optimizing injection strategies. Among various approaches, seismic monitoring has proven to be one of the most powerful tools for detecting subtle changes associated with subsurface fluid~\citep{chadwick2009latest, roach2018evolution, pevzner2021seismic}. Full-waveform monitoring (FWM), also known as time-lapse full-waveform inversion, is an advanced seismic monitoring technique that inverts the complete recorded wavefield to resolve spatial variations in the subsurface at high resolution~\citep{asnaashari2015time, maharramov2016time}. This method has been successfully applied in various monitoring contexts, including CO$_2$ injection~\citep{queisser2013full, egorov2017time}, hydrocarbon production~\citep{hicks2016time}, and near-surface aquifer recharge~\citep{li2025fiber}.

% Uncertainty quantification in FWM
Despite its promise, conventional FWM is typically formulated as a deterministic inverse problem that seeks a single model minimizing the waveform misfit~\citep{tarantola2005inverse}. Although such approaches can achieve adequate data fitting, they are vulnerable to the non-uniqueness inherent in ill-posed inverse problems and provide no measure of uncertainty. Under the sparse seismic acquisition typical of cost-constrained GCS operations~\citep{isaenkov2021automated}, deterministic inversions may produce multiple, equally plausible models that fit the data yet imply different CO$_2$ plume geometries and saturation distributions, resulting in ambiguous interpretations. They also tend to yield unstable or unreliable results when significant noise is present, especially under very sparse survey configurations. 

A Bayesian formulation of the inverse problem offers a statistical framework with posterior sampling~\citep{sambridge2002monte}. In this framework, the prior encodes geological or physical constraints, and the likelihood quantifies the mismatch between observed and modeled seismic waveforms. The resulting posterior distribution provides probabilistic estimates of subsurface properties. Various Monte Carlo~\citep{zhao2021gradient} and variational inference~\citep{zhang2020seismic} strategies have been applied to explore the posterior distribution. However, conventional Markov Chain Monte Carlo (MCMC) methods often suffer from slow convergence and poor scalability as model dimensionality increases~\citep{neal2011mcmc}. Recently, Hamiltonian Monte Carlo (HMC) has emerged as an appealing sampling method~\citep{gebraad2020bayesian, zunino2023hmclab}, as it leverages gradients of the log-posterior to propose distant, weakly correlated samples that efficiently traverse target probability landscape while maintaining asymptotic accuracy~\citep{neal2011mcmc}. These properties make HMC particularly promising for quantifying uncertainty in FWM of CO$_2$ injection.

% Informed priors is important in Bayesian FWM
While sampling strategies such as HMC mitigate some computational challenges of exploring high-dimensional posterior distributions, the efficiency of Bayesian inference also depends on the construction of informative, physics-consistent priors. Simple uniform priors or Gaussian assumptions of the model space cannot capture reservoir heterogeneity, and the complex migration patterns of a CO$_2$ plume. Previous studies have shown that incorporating priors constrained by geological structure~\citep{zhu2016bayesian}, rock physics relationships~\citep{li2016integrated, hu2023feasibility, mardan2023monitoring}, or petrophysical parameters~\citep{zunino2015monte, zhang2018multiparameter, aragao2020elastic} can improve seismic inversion performance. Using deep learning strategies to assimilate prior information in seismic inversion has also shown promising results~\citep{fang2020deep, zhu2022integrating, taufik2024learned}. In the context of seismic CO$_2$ monitoring, an effective prior is expected to honor permeability-porosity distributions informed by geostatistical information, the fluid-flow physics governing CO$_2$ plume migration, and the rock physics relationships linking fluid saturation to seismic velocity perturbations. Integrating these diverse sources of information into a single Bayesian FWM framework provides a multi-physics inference scheme that reduces the inversion null space by assimilating both seismic and non-seismic constraints~\citep{li2020coupled, yin2024time}.

% Proposed Methodology
Based on these considerations, we develop a Bayesian FWM framework that incorporates informative priors derived from fluid-flow simulations, represented compactly through a deep-learning-based generative algorithm. We first construct prior models using ensembles of geomodel realizations and fluid-flow simulations, which capture realistic patterns of CO$_2$ saturation variability. To represent and sample from these high-dimensional priors more efficiently, we adopt a variational autoencoder (VAE), a deep generative model that learns a low-dimensional representation of the saturation fields while preserving the multi-modal and non-Gaussian statistics of the original ensembles~\citep{kingma2013auto, laloy2017inversion, jiang2024history}. Compared with traditional linear dimensionality reduction methods such as truncated singular value decomposition or basis-function expansions~\citep{li2024time}, the VAE better captures the complex nonlinear spatial structures of CO$_2$ saturation and also supports differentiable sampling in its latent space~\citep[e.g.,][]{lopez2021deep}. The generated saturation fields are then converted into the P-wave velocity changes for wave-equation modeling via rock physics mapping~\citep[e.g.,][]{mavko2020rock, hu2023feasibility}. HMC sampling is performed directly in the latent space of the trained VAE, where gradient-based exploration and the learned generative prior jointly enable computationally efficient posterior sampling. Finally, we obtain posterior CO$_2$ saturation fields with quantified uncertainties.

% Outlines
This paper is organized as follows. We first introduce the proposed methodology, including geomodeling, fluid-flow simulation, physics-informed prior construction, VAE-based latent representation, rock physics relationships, and the formulation of HMC sampling. We then demonstrate the framework through numerical experiments using 2-D acoustic wave simulations designed to evaluate performance under different monitoring scenarios. This is followed by a discussion of broader implications, including prior design, sampling strategies, limitations, and potential extensions.

\section{Methodology}
\begin{figure}
\centering
\includegraphics[width=0.95\textwidth]{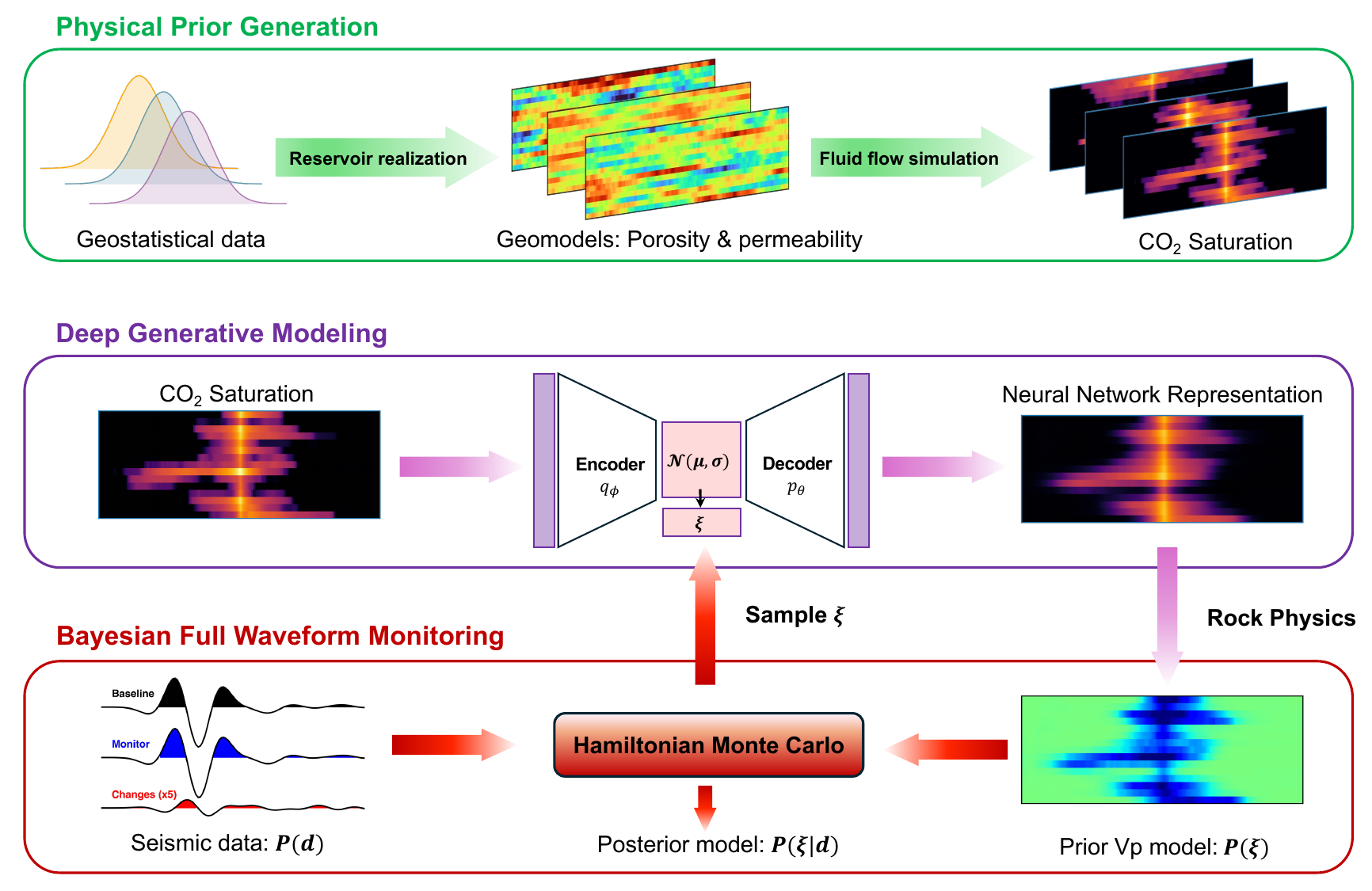}
\caption{Workflow of the proposed Bayesian FWM framework. Reservoir realizations are generated using geostatistical models and fluid-flow simulations to produce CO$_2$ saturation fields. A VAE learns a low-dimensional manifold of these fields to construct a generative prior, which is mapped to seismic velocities through rock physics modeling. HMC sampling in the latent space infers posterior distributions conditioned on time-lapse seismic data.}
\label{fig:workflow}
\end{figure}

This section presents the proposed probabilistic FWM framework, which integrates geological modeling, fluid-flow simulation, generative deep learning, and HMC sampling into a single inference workflow for CO$_2$ plume monitoring. As outlined in Figure~\ref{fig:workflow}, the proposed workflow constructs geostatistical- and physics-informed priors of the CO$_2$ plume, represents them in a compact latent space using a generative strategy, and conditions these priors on time-lapse seismic data to infer posterior distributions of saturation changes. Here, rock physics modeling maps saturation fields to seismic velocity models for wavefield simulation. The HMC sampler is used to perform gradient-based posterior exploration of the latent space. 

\subsection{Geomodels and Fluid-Flow Simulation}
We begin by generating reservoir realizations using geostatistical information. These realizations are then used in fluid-flow simulations under CO$_2$ injection to obtain saturation fields. The flow simulations solve the mass conservation equations coupled with Darcy’s law to model the migration of the injected supercritical CO$_2$. Specifically, we adopt the 3-D geomodel following \citet{wang2025deep}. The computational domain spans $109~\mathrm{km} \times 109~\mathrm{km} \times 70~\mathrm{m}$ and contains a central storage aquifer of size $896~\mathrm{m} \times 896~\mathrm{m} \times 70~\mathrm{m}$, and a large surrounding region. The storage aquifer is discretized into $128 \times 128 \times 35$ cells, while the full domain is discretized into $148 \times 148 \times 35$ cells, with the surrounding region represented at a much coarser resolution. 

Uncertainty in the storage aquifer properties is represented through a set of meta-parameters and associated (random) realizations. The meta-parameters include the mean ($\mu_{\log k}$) and standard deviation ($\sigma_{\log k}$) of log-permeability, the anisotropy ratio ($a_r$), and the permeability-porosity relationship parameters ($d$ and $c$). The anisotropy ratio represents the relationship between vertical and horizontal permeability, defined as $a_r=k_v/k_h$, where $k_v$ and $k_h$ represent vertical and horizontal permeability, respectively. The parameters $d$ and $c$ define the permeability-porosity relationship via $\phi=d\cdot \log k+c$. The parameter ranges are summarized in Table~\ref{table:meta_para}. They are assumed to be obtained from geological data, such as borehole logs. For any set of meta-parameters, an arbitrarily large number of realizations can be generated. In this work, SGeMS~\citep{remy2009applied} is applied, with a spherical variogram of correlation lengths $l_x=l_y=280$~m and $l_z=7$~m, for realization generation. See \citet{wang2025deep} for further details on geomodel construction.

\begin{table}[htbp]
\setlength{\belowcaptionskip}{0.25cm}
\centering
\caption{meta-parameters and their ranges. $U[a,b]$ denotes a uniform distribution between $a$ and $b$.}
\makebox[\textwidth][c]{
\begin{tabular}{lc}
\hline
\bf{Meta-parameter} & \bf{Range} \\
\hline
Mean of log-permeability, $\mu_{\log k}$ & $U[2, 6]$ \\
Standard deviation of log-permeability, $\sigma_{\log k}$ & $U[1.0, 2.5]$ \\
Log of permeability anisotropy ratio, $\log_{10}a_r$ & $U[-2, 0]$ \\
Parameter $d$ & $U[0.02, 0.05]$ \\
Parameter $c$ & $U[0.05, 0.12]$ \\
\hline
\end{tabular}
}
\label{table:meta_para}
\end{table}

A vertical well with a constant injection rate of 0.5~kmt~CO$_2$/year is placed at the center of the storage aquifer. The injector penetrates all layers, and the total injection time is one year, which is divided into 30 time steps. Flow simulations of CO$_2$ migration during the injection process for different geomodels are performed using the numerical simulator GEOS~\citep{settgast2024geos}. In total, 4,000 geomodel realizations are used to generate the 3-D saturation fields. Because our seismic monitoring framework is 2-D, two vertical 2-D sections, one in the $x$-direction and one in the $y$-direction (both intersecting the well), are extracted from each 3-D saturation field. The resulting 8,000 2-D saturation realizations represent the priors used in the rest of this study. These realizations characterize a wide range of spatial and temporal variability in CO$_2$ flow behavior.

\subsection{Generative Prior via Variational Autoencoder}
To handle high-dimensional prior models constructed in the previous section, we represent the CO$_2$ saturation fields in a compact latent space using a VAE~\citep{kingma2013auto}. Direct sampling in the original model space is computationally expensive due to its high dimensionality and complex distribution pattern. This is the case even though HMC scales more favorably than traditional Metropolis-Hastings algorithms, with independent-sample efficiency scaling as $\mathcal{O}(n^{5/4})$ rather than $\mathcal{O}(n^2)$, where $n$ is the dimension of the model space~\citep{neal2011mcmc}. This motivates the use of a low-dimensional, differentiable (supporting gradient-based sampling), and expressive prior that preserves the key characteristics of the original saturation realizations.

We employ a VAE to construct a generative prior for CO$_2$ saturation fields within the storage aquifer region. A VAE consists of two neural networks, including an encoder and a decoder. Given an ensemble of saturation models $\{\boldsymbol{S}_i\}_{i=1}^N \subset \mathbb{R}^M$ (where $N$ denotes the total number of prior realizations and $M$ denotes the dimension of the original models), the encoder projects each model $\boldsymbol{S}_i$ to the parameters of a Gaussian distribution in the latent space, expressed as follows:
\begin{equation}
q_{\phi}(\boldsymbol{\xi}\,|\,\boldsymbol{S}_i) = 
\mathcal{N}(\boldsymbol{\mu}_i,\ \mathrm{diag}(\boldsymbol{\sigma}_i^2)),
\label{eq:encoder}
\end{equation}
where $\boldsymbol{\mu}_i$ and $\boldsymbol{\sigma}_i$ are the encoder outputs, and $q_{\phi}$ denotes the encoder with trainable parameters. A latent vector $\boldsymbol{\xi} \in \mathbb{R}^K$, where $K$ is the dimensionality of the latent space (typically much smaller than the model dimensionality of the original CO$_2$ saturation fields), is sampled from this distribution and passed through the decoder $p_{\theta}$:
\begin{equation}
\hat{\boldsymbol{S}}_i = p_{\theta}(\boldsymbol{\xi}_i),
\label{eq:decoder}
\end{equation}
to produce a reconstructed saturation field $\hat{\boldsymbol{S}}_i$.

The VAE is trained by maximizing the evidence lower bound (ELBO)~\citep{kingma2013auto}, where $\phi$ and $\theta$ represent the trainable parameters of the encoder and decoder. This training balances two objectives: (1) accurately reconstructing the input saturation fields (measured using mean squared error), and (2) regularizing the latent variables towards a standard normal distribution through a Kullback-Leibler (KL) penalty. The KL regularization promotes a smooth and continuous latent space, which is important for stable generative sampling and subsequent gradient-based Bayesian inversion.

We trained a VAE on the ensemble of 8,000 2-D CO$_2$ plume realizations. Each 2-D sample was resampled onto a $179\times 15$ numerical grid (corresponding to a reservoir domain of $896~\mathrm{m} \times 70~\mathrm{m}$ with $\sim$5~m spatial resolution). Of these samples, 7,000 were used for training and 1,000 for validation. The network architecture is provided in Supporting Table~S1. Specifically, we adopt the $\textit{tanh}$ activation function at the decoder output to constrain saturation values between 0 and 1. A latent dimension of 64 was selected to capture the dominant variability in the saturation fields. The KL divergence weight was set to $1.5 \times 10^{-5}$ to balance reconstruction fidelity and generation stability, which provides a reasonable compromise between the two objectives based on empirical testing.

Once trained, new CO$_2$ saturation realizations can be generated by first sampling latent vectors from a standard normal distribution, $\boldsymbol{\xi} \sim \mathcal{N}(\boldsymbol{0}, \boldsymbol{I})$, and then decoding them using Eq.~\ref{eq:decoder}. This approach maps the original high-dimensional, non-Gaussian model parameters into a lower-dimensional Gaussian latent space. In our case, the original saturation fields contain $2{,}685$ $(179 \times 15)$ unknowns in the targeted storage aquifer layer (as outlined in Figure~\ref{fig:model}a), whereas a latent dimension of 64 is sufficient to capture their dominant variability. This dimensionality reduction by a factor of 40 substantially decreases the number of parameters to be inferred. Moreover, the original 2,685 grid parameters are spatially correlated (i.e., nearby cells vary together, resulting in an elongated, anisotropic parameter distribution), which makes the posterior distribution difficult to explore. The VAE maps the original model space into a lower-dimensional latent space that follows an approximately standard normal distribution. This reparameterization improves the posterior structure, enabling more efficient HMC sampling by reducing autocorrelation and increasing the effective sample size per gradient evaluation.

\subsection{Rock Physics Modeling}
The trained VAE provides an effective prior representation of CO$_2$ saturation fields. To map these saturation realizations into seismic properties required for waveform modeling, we next incorporate rock physics relations that link fluid substitution to seismic parameter changes. In this workflow, brine properties (the initial pore fluid) are computed following~\citet{batzle1992seismic} using reservoir conditions of 60$^\circ$C, 14~MPa, and a salinity of 0.011 (mass fraction). Supercritical CO$_2$ properties are obtained from the equation of state~\citep{span1996new} at the same temperature and pressure. For a given CO$_2$ saturation $\boldsymbol{S}$, the effective fluid bulk modulus is evaluated using the Brie empirical mixing law \citep{mavko2020rock} with Brie's coefficient $e=2$:
\begin{equation}
K_f = (K_w - K_{\mathrm{CO_2}})\ (1 - \boldsymbol{S})^{\,e} + K_{\mathrm{CO_2}},
\label{eq:brie}
\end{equation}
where $K_w$ and $K_{\mathrm{CO_2}}$ are the bulk moduli of brine and CO$_2$, respectively. The mixed fluid density is computed using the standard volume-fraction mixing law.

The resulting effective fluid properties are incorporated into the solid frame via Gassmann fluid substitution~\citep{mavko2020rock}, with the saturated bulk modulus given by:
\begin{equation}
K_{\mathrm{sat}} 
= K_{\mathrm{dry}} 
+ \frac{\left(1 - \dfrac{K_{\mathrm{dry}}}{K_{\mathrm{min}}}\right)^2}
       {\dfrac{\phi}{K_f} + \dfrac{1-\phi}{K_{\mathrm{min}}} - \dfrac{K_{\mathrm{dry}}}{K_{\mathrm{min}}^2}},
\label{eq:rock}
\end{equation}
where $K_{\mathrm{dry}}$ is the dry-frame bulk modulus, $K_{\mathrm{min}}$ is the mineral bulk modulus, and $\phi$ is the porosity. Saturated density is obtained by mixing the mineral and fluid densities in proportion to their volume fractions, while the shear modulus is assumed to remain unchanged during fluid substitution. With the updated moduli and density, the P-wave velocity is computed as follows:
\begin{equation}
\boldsymbol{v}_p = \sqrt{\frac{K_{\mathrm{sat}} + \frac{4}{3}\mu}{\rho_{\mathrm{sat}}}},
\label{eq:vp_update}
\end{equation}
which is used in acoustic wave-equation simulations to model the time-lapse seismic data in the Bayesian monitoring framework.

Although a deterministic rock physics model is adopted here, the workflow readily accommodates stochastic extensions by assigning prior distributions to uncertain quantities such as mineral moduli, porosity, or the Brie's coefficient $e$. This would allow rock physics uncertainty to be propagated directly into the seismic velocity models. Depending on reservoir conditions and monitoring objectives, alternative formulations, such as patchy saturation models or elastic frame alterations, may also be considered. The model used here is intended to demonstrate the integration of fluid-flow priors with seismic forward modeling.

\subsection{Bayesian Inference in Latent Space}
With the generative prior in place, Bayesian inference can be performed directly in the VAE latent space. The goal is to obtain the posterior probability distribution of the latent variable $\boldsymbol{\xi}$ conditioned on the observed time-lapse seismic data. In this latent formulation, the prior probability density function (PDF) for $\boldsymbol{\xi} \in \mathbb{R}^K$ is a standard multivariate normal distribution:
\begin{equation}
p(\boldsymbol{\xi}) = \mathcal{N}(\boldsymbol{0}, \boldsymbol{I}),
\label{eq:prior}
\end{equation}
which encodes the learned variability of CO$_2$ saturation fields through the trained VAE. Sampling from this distribution and decoding via Eq.~\ref{eq:decoder} produces realistic CO$_2$ saturation fields.

Each saturation field $\boldsymbol{S}$ is mapped to a P-wave velocity model $\boldsymbol{v}_p$ using the rock physics model in Eq.~\ref{eq:rock} and Eq.~\ref{eq:vp_update} for wave-equation simulation. The synthetic seismic data $\boldsymbol{d}$ are generated by solving the 2-D acoustic wave equation expressed as follows:
\begin{equation}
\frac{\partial^2 u(\boldsymbol{x}, t)}{\partial t^2}
-\nabla \cdot \left(\boldsymbol{v}_p^2(\boldsymbol{x})\, \nabla u(\boldsymbol{x}, t)\right)
= s(\boldsymbol{x}_s, t),
\label{eq:wave_eq}
\end{equation}
where $u(\boldsymbol{x}, t)$ denotes the seismic wavefield and $s(\boldsymbol{x}_s, t)$ denotes the seismic source. Synthetic data $\boldsymbol{d}$ correspond to the wavefield $u(\boldsymbol{x}, t)$ sampled at geophone positions. Wavefield simulations are performed using a GPU-accelerated wave-equation propagator~\citep{richardson_2023_8381177}.

The likelihood function is defined using the $\ell_2$ norm to measure the discrepancy between observed seismic data $\boldsymbol{d}^{\mathrm{obs}}$ and synthetic data $\boldsymbol{d}$ as follows:
\begin{equation}
p\left(\boldsymbol{d}^{\mathrm{obs}} \mid \boldsymbol{\xi} \right) 
\propto \exp\left( 
    -\frac{1}{2\sigma^2 T} 
    \left\| \boldsymbol{d}(\boldsymbol{\xi}) - \boldsymbol{d}^{\mathrm{obs}} \right\|^2 
\right),
\label{eq:likelihood}
\end{equation}
where $\sigma^2$ denotes the measurement-error variance (assumed identical for all observations), and $T$ is the simulated-annealing temperature that controls the breadth of posterior exploration~\citep{kirkpatrick1983optimization}. Bayes’ theorem combines the prior (Eq.~\ref{eq:prior}) and likelihood (Eq.~\ref{eq:likelihood}) to yield the posterior distribution of the latent variable:
\begin{equation}
p(\boldsymbol{\xi} \mid \boldsymbol{d}^{\mathrm{obs}}) = \frac{p(\boldsymbol{d}^{\mathrm{obs}} \mid \boldsymbol{\xi}) \cdot p(\boldsymbol{\xi})}{p(\boldsymbol{d}^{\mathrm{obs}})},
\label{eq:bayes_rule}
\end{equation}
where the evidence $p(\boldsymbol{d}^{\mathrm{obs}})$ is a normalization factor and is not considered when sampling from the posterior. Sampling from the posterior over $\boldsymbol{\xi}$ yields models consistent with both seismic observations and the prior informed by geomodels and fluid-flow physics. By decoding these latent-space posterior samples (Eq.\ref{eq:decoder}), we then obtain the posterior distribution of the CO$_2$ saturation field in the physical domain. 

\subsection{Hamiltonian Monte Carlo Sampling}
To sample from the posterior distribution $p(\boldsymbol{\xi} \mid \boldsymbol{d}^{\mathrm{obs}})$, we employ the gradient-based HMC sampler, which leverages concepts from classical mechanics to explore high-dimensional probability landscape efficiently. In HMC, the model parameters, namely the latent variables ($\boldsymbol{\xi} \in \mathbb{R}^K$), are conceptualized as the location of a hypothetical physical particle in the $K$-dimensional model space. This hypothetical particle moves governed by the potential energy $U$, which is defined as follows:
\begin{equation}
    U(\boldsymbol{\xi}) = - \ln p(\boldsymbol{\xi}).
\end{equation}
To describe the state of the Markov chain, an auxiliary momentum variable $\boldsymbol{p} \in \mathbb{R}^K$ is introduced (note that $\boldsymbol{p}$ denotes momentum, whereas probability is denoted by $p$), along with a generalized mass matrix $\boldsymbol{M}$ of dimension $K \times K$. The kinetic energy is then defined as:
\begin{equation}
    K(\boldsymbol{p}) = \frac{1}{2} \boldsymbol{p}^T \boldsymbol{M}^{-1} \boldsymbol{p}.
\label{eq:kinetic}
\end{equation}
Here, the momentum $\boldsymbol{p}$ is randomly sampled from a multivariate Gaussian distribution with covariance matrix $\boldsymbol{M}$. The mass matrix governs the scaling and coupling of momentum components, allowing the HMC sampler to move across different energy levels and traverse the posterior landscape. The system Hamiltonian, which combines potential and kinetic energies, is expressed as:
\begin{equation}
    H(\boldsymbol{\xi}, \boldsymbol{p}) = U(\boldsymbol{\xi}) + K(\boldsymbol{p}).
\end{equation}

To sample from the distribution, one begins with a model $\boldsymbol{\xi}_0$ drawn from the prior distribution $p(\boldsymbol{\xi})$ and a randomly selected initial momentum $\boldsymbol{p}_0$. The position of the hypothetical particle is evolved in an artificial time variable $\tau$ according to Hamilton’s equations:
\begin{equation}
\frac{d \boldsymbol{\xi}}{d \tau}=\frac{\partial H}{\partial \boldsymbol{p}} = \boldsymbol{M}^{-1} \boldsymbol{p}, \quad \frac{d \boldsymbol{p}}{d \tau}=-\frac{\partial H}{\partial \boldsymbol{\xi}}= - \frac{\partial U}{\partial \boldsymbol{\xi}}.
\end{equation}
This system is then integrated using a symplectic leapfrog integrator, which yields a proposal $(\tilde{\boldsymbol{\xi}}, \tilde{\boldsymbol{p}})$~\citep{neal2011mcmc}. Due to numerical errors in the leapfrog integration, the Hamiltonian is not exactly conserved. These small deviations may cause the proposal distribution to slightly differ from the true target distribution. To correct for this and ensure that the Markov chain samples from the correct posterior, the Metropolis-Hastings criterion~\citep{metropolis1953equation} is applied at the end of each trajectory as follows:
\begin{equation}
p_{\mathrm{accept}} = \min\left(1, \exp \left[ H(\boldsymbol{\xi}, \boldsymbol{p}) - H(\tilde{\boldsymbol{\xi}}, \tilde{\boldsymbol{p}}) \right] \right).
\label{eq:mh_criterion}
\end{equation}
If the proposal is accepted, the new state $(\tilde{\boldsymbol{\xi}}, \tilde{\boldsymbol{p}})$ is added to the chain; otherwise, the previous state is retained. This mechanism allows HMC to effectively sample the joint distribution of momentum-model phase space, and the distribution over the model parameters $p(\boldsymbol{\xi})$ is obtained by marginalizing out the momentum component.

In the HMC framework, the model space corresponds to the latent space of the variable $\boldsymbol{\xi}$, and the posterior potential $U(\boldsymbol{\xi})$ is related to the misfit of seismic waveform data. The required gradient $\partial U / \partial \boldsymbol{\xi}$ is computed via the chain rule:
\begin{equation}
\frac{\partial U}{\partial \boldsymbol{\xi}} = 
\frac{\partial U}{\partial \boldsymbol{d}} \cdot  
\frac{\partial \boldsymbol{d}}{\partial \boldsymbol{v}_{p}} \cdot
\frac{\partial \boldsymbol{v}_{p}}{\partial \hat{\boldsymbol{S}}} \cdot 
\frac{\partial \hat{\boldsymbol{S}}}{\partial \boldsymbol{\xi}},
\label{eq:chain_rule}
\end{equation}
where ${\partial U}/{\partial \boldsymbol{d}}$ denotes the sensitivity of the posterior potential to the seismic data residual, ${\partial \boldsymbol{d}}/{\partial \boldsymbol{v}_{p}}$ denotes the adjoint-state gradient of the data misfit with respect to the P-wave velocity, ${\partial \boldsymbol{v}_{p}}/{\partial \hat{\boldsymbol{S}}}$ denotes the sensitivity of rock physics mapping to CO$_2$ saturation, and ${\partial \hat{\boldsymbol{S}}}/{\partial \boldsymbol{\xi}}$ denotes the Jacobian of the VAE decoder relating the saturation field to the latent variables.

\section{Numerical Experiments}
Building on the methodology described above, we demonstrate the proposed inference framework through a series of numerical experiments. These experiments are designed to assess the generative prior via VAE and the latent-space HMC sampling in monitoring CO$_2$ saturation using time-lapse seismic data.

\begin{figure}[H]
\centering
\includegraphics[width=1.0\textwidth]{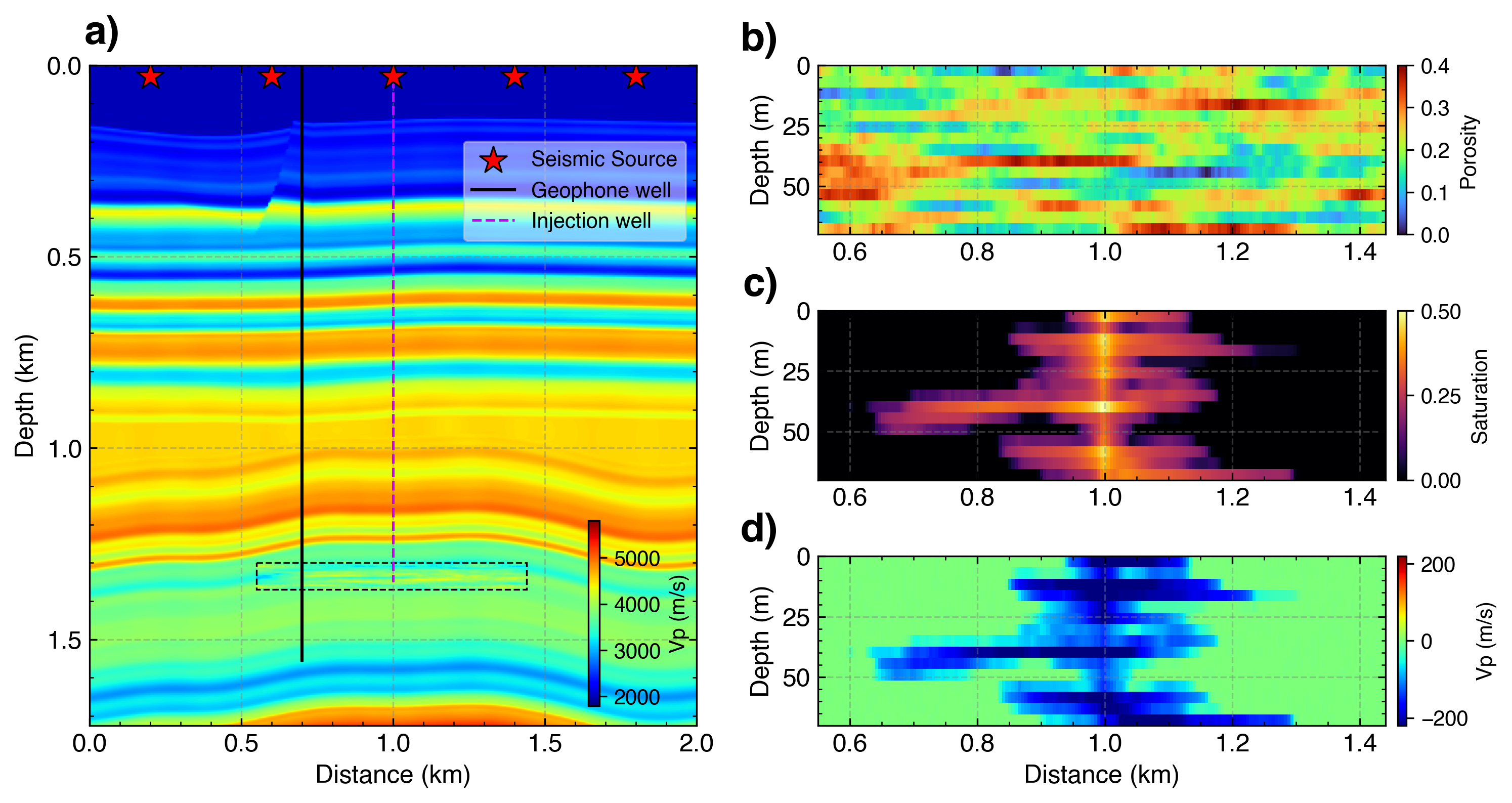}
\caption{
(a) Baseline P-wave velocity model for the geological CO$_2$ injection. The dashed box marks the target reservoir, spanning lateral positions of 550-1,440~m and depths of 1,300-1,370~m. The purple dashed line indicates the injection well, and the black solid line denotes the observational geophone well. Red stars indicate the five surface seismic sources. The selected monitoring geomodel realization and its associated porosity field (b), CO$_2$ saturation distribution at the 10th month of injection (c), and the resulting P-wave velocity change (d) are also shown. Note the zero depth value in (b), (c) and (d) corresponds to a depth of 1,300~m in (a).}
\label{fig:model}
\end{figure}

Figure~\ref{fig:model}a shows the synthetic baseline P-wave velocity model for the CO$_2$ storage setting, where the target storage aquifer (outlined by the box) spans lateral positions from 550 to 1,440 m and depths of 1,300 to 1,370 m. The full model has a grid size of $401 \times 346$, while the storage aquifer area corresponds to a $179 \times 15$ grid subset. The injection well is shown with a purple dashed line, and the observational well with geophones is shown with a black line. The observational well is set deeper than the storage aquifer to record transmission waves through the CO$_2$ plume. Five surface seismic sources are placed at 400 m intervals to form a vertical seismic profile survey. We use a 30 Hz central-frequency wavelet and record 1.2 s of data. One synthetic geomodel realization is selected as the target monitoring model for the subsequent numerical experiments. Figures~\ref{fig:model}b-d illustrate the porosity field, the simulated CO$_2$ saturation at the 10th month of injection, and the associated P-wave velocity changes.

\subsection{Generative Prior Evaluation}
\begin{figure}
\centering
\includegraphics[width=1.0\textwidth]{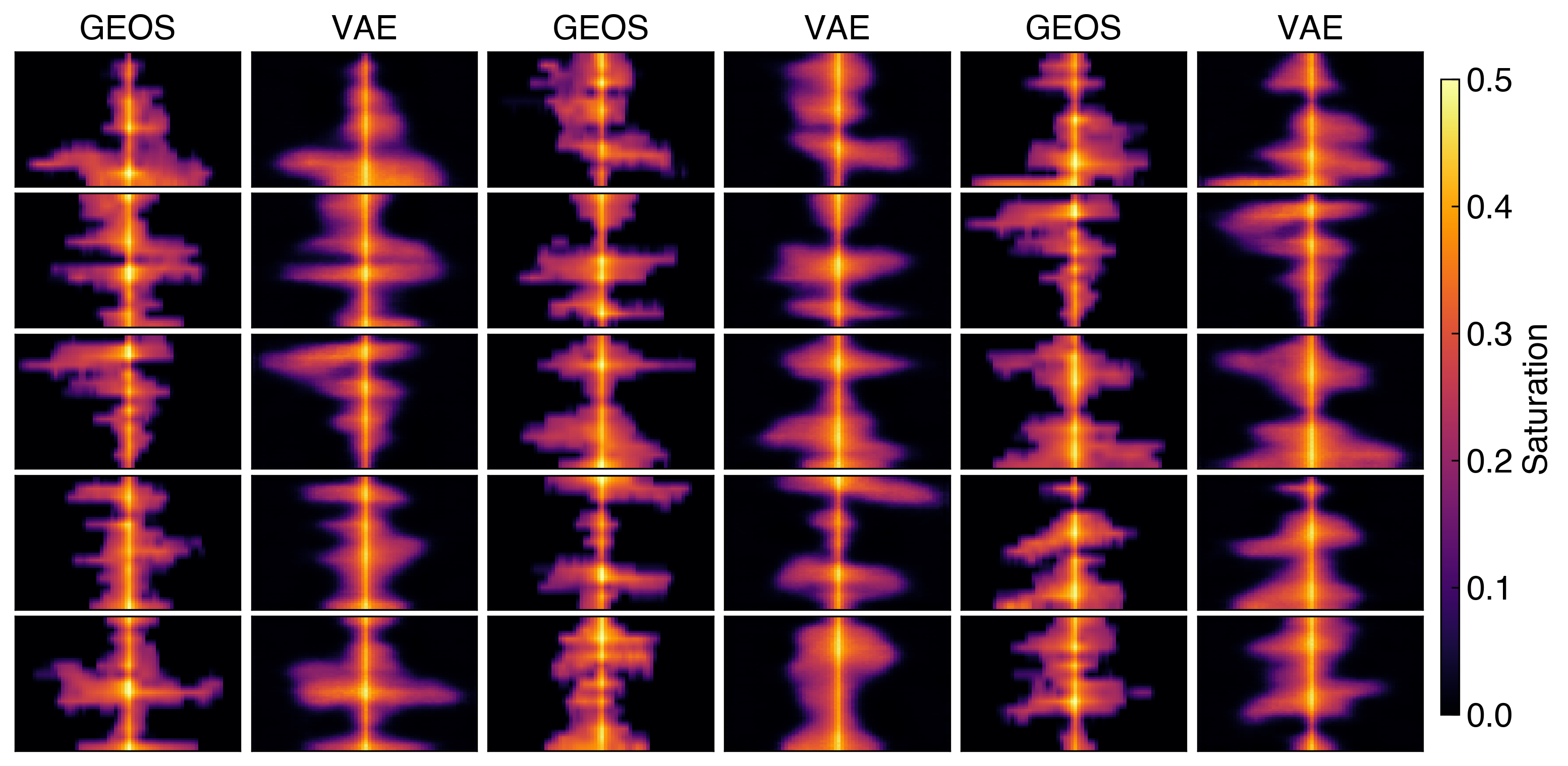}
\caption{Comparison between the original CO$_2$ saturation fields modeled with GEOS and their VAE reconstructions on the validation dataset. The spatial domain corresponds to the target reservoir interval outlined in Figure~\ref{fig:model}a, spanning depths of 1,300-1,370~m and lateral distance of 550-1,440~m. Column~2 represents VAE reconstructions of the GEOS results in column~1, column~4 shows reconstructions of the results in column~3, and column~6 provides reconstructions of the results in column~5.}
\label{fig:prior_reconstruction}
\end{figure}

Before performing Bayesian inference, we first assess the quality of the generative prior by examining whether the trained VAE model can produce realistic CO$_2$ plume realizations that honor the statistics of the original prior ensemble. The VAE model used in the subsequent section was trained with a learning rate of 0.001 (see the convergence curve in Supporting Figure~S1), with its architecture, training procedure, and data preprocessing detailed in the Methodology section. Figure~\ref{fig:prior_reconstruction} compares the original CO$_2$ saturation fields simulated with GEOS against their VAE reconstructions on the validation dataset. The reconstructions preserve key spatial structures, including plume geometry, connectivity, and saturation values. Some smoothing near the plume boundaries is observed, which is typical of VAE models employing convolutional layers. These fine-scale differences lie below the detectable resolution of the data and thus do not significantly impact the subsequent inversion. The overall agreement indicates that the trained VAE successfully captures the plume shape. Additional examples generated by decoding randomly sampled latent vectors are provided in Supporting Figure~S2, showing the stability and quality of the generative results. Statistical consistency between the generated and original saturation priors is demonstrated in Supporting Figure~S3, where comparisons of the $P_{10}$, $P_{50}$, and $P_{90}$ percentiles suggest that the ensemble generated with the VAE closely reproduces the statistics of the realizations simulated with GEOS.

\begin{figure}
\centering
\includegraphics[width=1.0\textwidth]{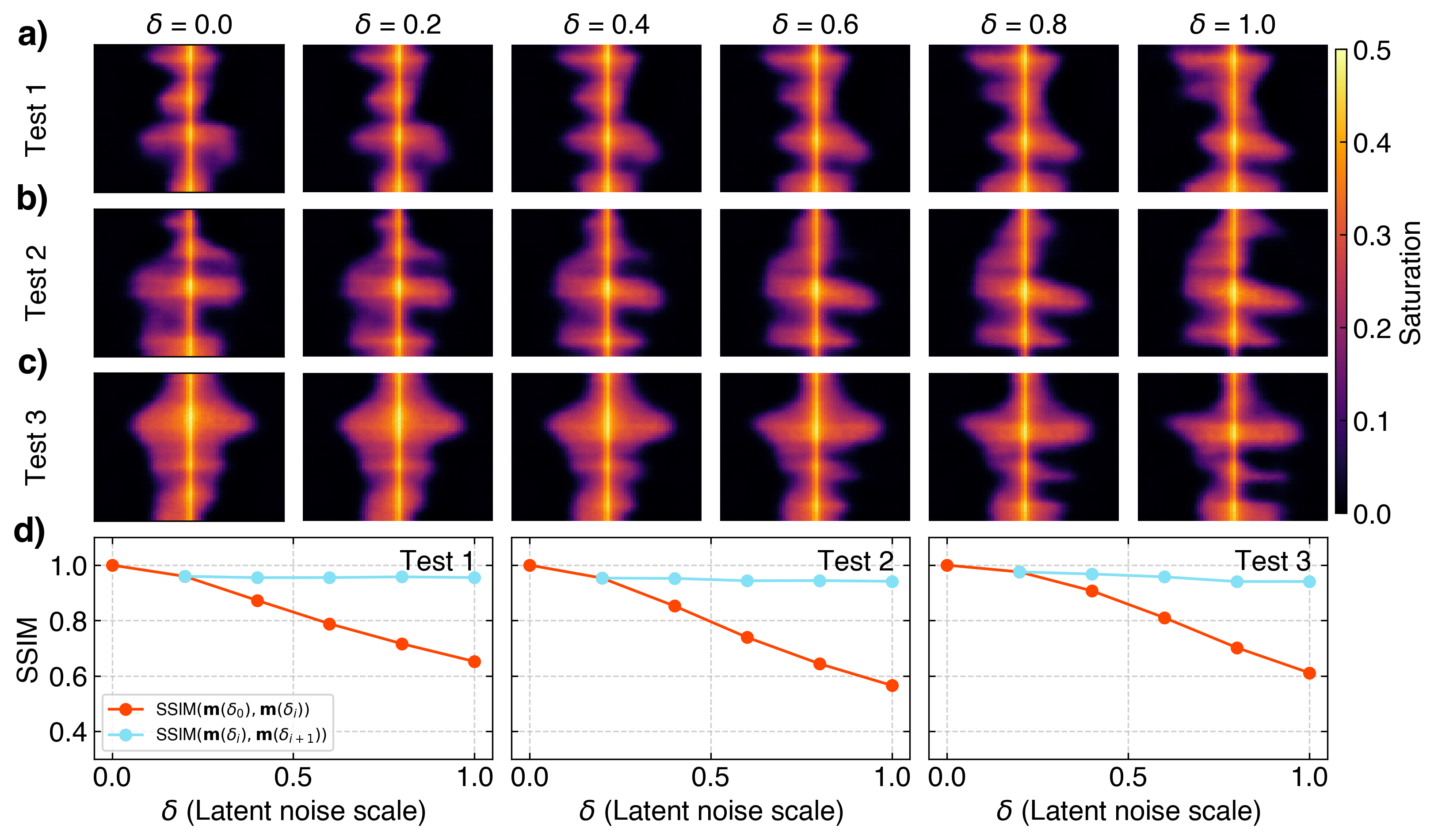}
\caption{Latent-space interpolation tests. (a-c) Saturation realizations generated at interpolation steps $\delta = 0.0$ to $1.0$ for three random latent vector pairs. (d) Sky-blue curves show similarity as measured by SSIM between consecutive interpolants, and orange curves show similarity between each interpolant and the end-member at $\delta = 0$.}
\label{fig:prior_latent_continuity}
\end{figure}

Since the HMC sampling requires gradients of the misfit function with respect to VAE latent variables (see Eq.~\ref{eq:chain_rule}), it is necessary to verify the continuity and structure of the learned latent space. We therefore conduct interpolation experiments between randomly selected latent vector pairs $\boldsymbol{\xi}^1$ and $\boldsymbol{\xi}^2$, following~\cite{di2025latent}. For each pair, intermediate latent vectors are obtained via linear interpolation:
\begin{equation}
\boldsymbol{\xi}(\delta) = (1-\delta)\boldsymbol{\xi}^1 + \delta \boldsymbol{\xi}^2, \quad \delta \in [0, 1].
\label{eq:latent_interp}
\end{equation}
Figures~\ref{fig:prior_latent_continuity}a-c show three sets of latent-space interpolation tests using interpolation steps of $\delta = 0.0$, 0.2, 0.4, 0.6, 0.8, and 1.0. The decoded saturation fields evolve smoothly between the two end-member realizations ($\delta = 0.0$ and $\delta = 1.0$). Continuity is quantified using the Structural Similarity Index Measure (SSIM), as shown in Figure~\ref{fig:prior_latent_continuity}d. SSIM values between consecutive interpolants remain high (sky-blue curves), while similarity relative to the end-member decreases gradually (orange curve). These results confirm that the latent space of the trained VAE is well-behaved.

\subsection{Bayesian Inference Results}
% Inversion setup
Having validated the quality of the generative prior via VAE, we now present the Bayesian inference results obtained through HMC sampling in the latent space. For the HMC inference, we adopt the adaptive tuning of the step size during 200 warm-up iterations to maintain a target acceptance rate of 0.65~\citep{bingham2019pyro}. In the HMC sampler, each trajectory consisted of 10 leapfrog steps, and the mass matrix (see Eq.~\ref{eq:kinetic}) was kept fixed without adaptive updates. After the burn-in phase, 40,000 posterior samples were generated. The results presented here use a temperature of 0.05, which yields neither an overly diffuse nor an overly concentrated posterior distribution.

\begin{figure}
\centering
\includegraphics[width=0.9\textwidth]{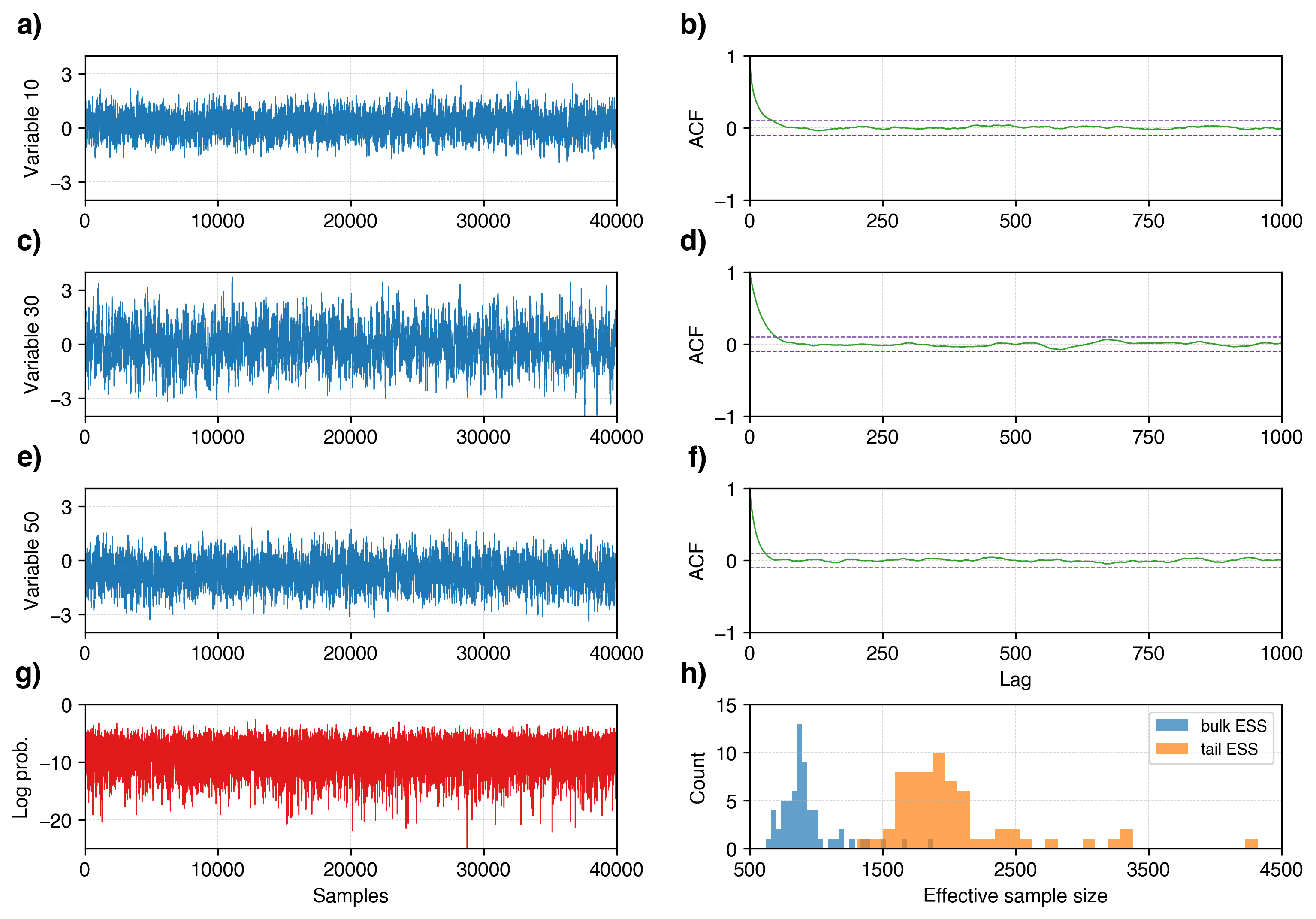}
\caption{Sampling diagnostics of the HMC results after the burn-in phase. (a,c,e) Trace plots for latent variables of dimensions 10, 30, and 50, respectively. (b,d,f) Corresponding autocorrelation functions (ACF) truncated at lag 1{,}000, where the dashed lines denote the $\pm 0.1$ bounds. (g) Negative log-probability across samples. (h) Distribution of bulk and tail effective sample sizes (ESS) across all 64 latent variables.}
\label{fig:hmc_trace}
\end{figure}

% Sampling Diagnostics
Convergence of the Markov chain was evaluated using several standard diagnostics~\citep{robert2004diagnosing}. The sampling traces of three latent dimensions (indices 10, 30, and 50 out of 64) are shown in Figures~\ref{fig:hmc_trace}a, c, and e. These trace plots show stable exploration around means with no divergence or drift, indicating good mixing. The full set of posterior latent traces is provided in Supporting Figure~S4. The corresponding autocorrelation functions (ACF) (Figures~\ref{fig:hmc_trace}b, d, f) decay rapidly, crossing below the $\pm 0.1$ threshold (dashed lines) within short lags. The negative log probability (Figure~\ref{fig:hmc_trace}g) remains stable throughout sampling. Across all 64 latent dimensions, the bulk and tail effective sample sizes (ESS) exceed 619 and 1,311 (Figure~\ref{fig:hmc_trace}h) out of the total of 40,000 samples, respectively. This suggests accurate estimates of both central and tail posterior summaries. The Monte Carlo standard error (MCSE) of the posterior standard deviation (SD) is also small relative to the posterior spread, where MCSE(SD)/SD ranges from 1.6\% to 2.9\% (median 2.4\%). Overall, these diagnostics demonstrate appropriate sampling convergence and sufficiently accurate posterior summaries for the following uncertainty analysis.

\begin{figure}
\centering
\includegraphics[width=0.9\textwidth]{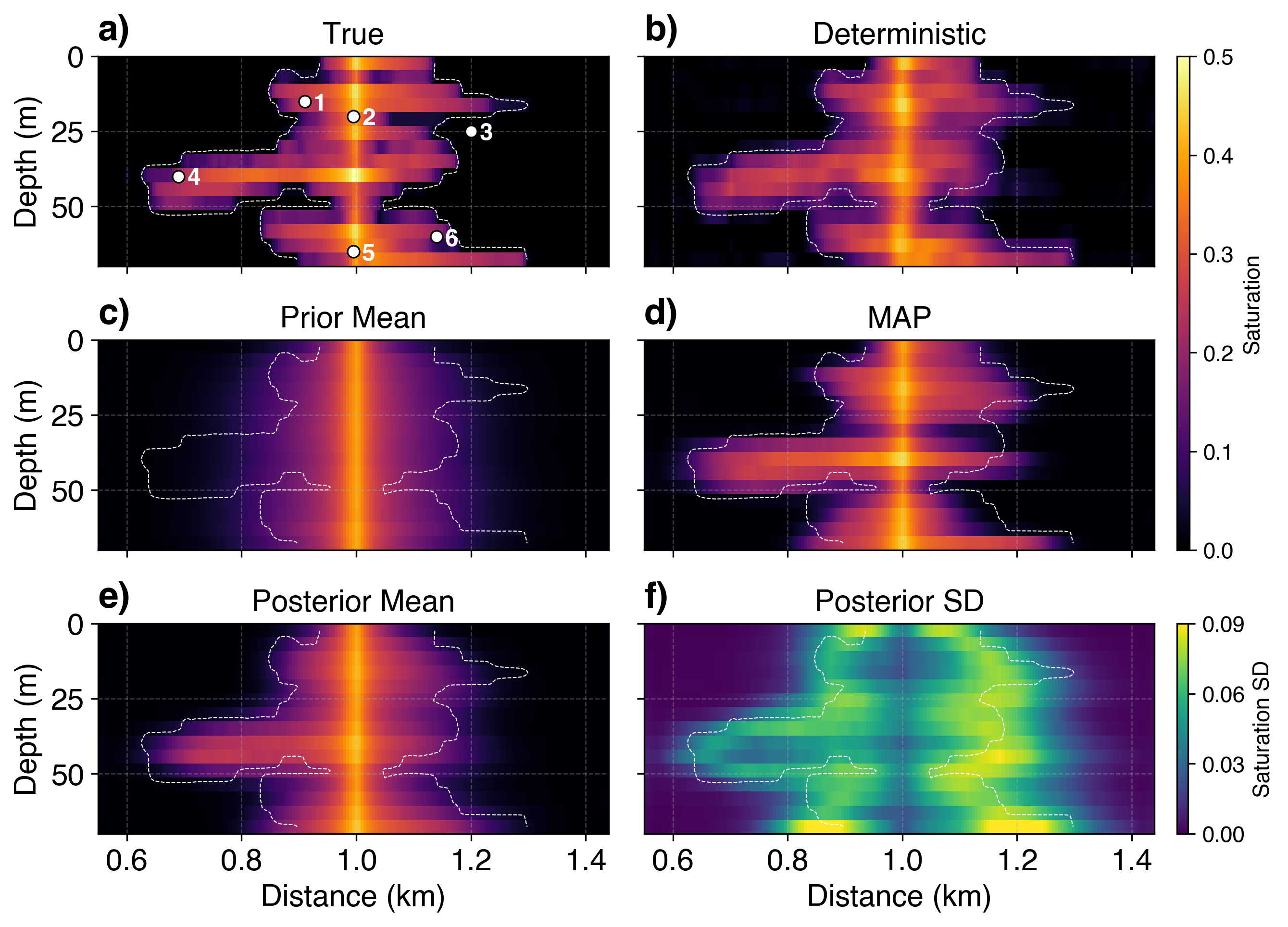}
\caption{
Bayesian monitoring results of CO$_2$ saturation.  
(a) True CO$_2$ plume.  
(b) Deterministic monitoring result ($l$-BFGS method).  
(c) Prior mean saturation (no seismic data assimilated).  
(d) Maximum a posteriori (MAP) estimate from the HMC posterior ensemble.  
(e) Posterior mean from the HMC posterior ensemble.  
(f) Posterior standard deviation (SD) from the HMC posterior ensemble.  
The white dashed line denotes the 0.03 saturation contour of the true model for reference. The locations labeled 1-6 are for subsequent posterior histogram analysis. Note the zero depth value corresponds to a depth of 1.3~km in Figure~\ref{fig:model}a.
}
\label{fig:hmc_result}
\end{figure}

% Posterior results
The Bayesian inference results for CO$_2$ saturation are provided in Figure~\ref{fig:hmc_result}, where the white dashed line marks the contour with a saturation value of 0.03 of the true model (Figure~\ref{fig:hmc_result}a). Figure~\ref{fig:hmc_result}c shows the mean of the saturation prior (without any seismic data assimilation). As expected, this result displays significant discrepancy with the true CO$_2$ plume. Figure~\ref{fig:hmc_result}b shows the deterministic monitoring result obtained using $l$-BFGS optimization method. The inversion adopts a target-oriented strategy restricted to the storage aquifer domain and assumes a uniform saturation prior bounded between 0 and 1. Differences from the Bayesian results are therefore expected due to the use of different priors and inversion strategies; however, when the seismic data are sufficiently informative, the deterministic result should be consistent with the maximum a posteriori (MAP) estimate of the Bayesian inversion.

As shown in Figure~\ref{fig:hmc_result}d, the MAP estimate recovers both the plume structure and saturation value, closely matching the true model. The posterior mean (Figure~\ref{fig:hmc_result}e) provides an averaged estimate of the CO$_2$ plume that smooths the sharp boundaries while preserving the overall geometry. The posterior standard deviation (Figure~\ref{fig:hmc_result}f) quantifies spatial uncertainty in the inferred saturation field, which shows low variability near the injection well and increased uncertainty around the right-hand plume margins. To further interpret the sampled posterior, we performed a clustering analysis using a k-means and medoid-based approach (see Supporting Figure~S5). The resulting medoids share consistent features, suggesting that the primary plume geometry is well captured. 

\begin{figure}
\centering
\includegraphics[width=1.0\textwidth]{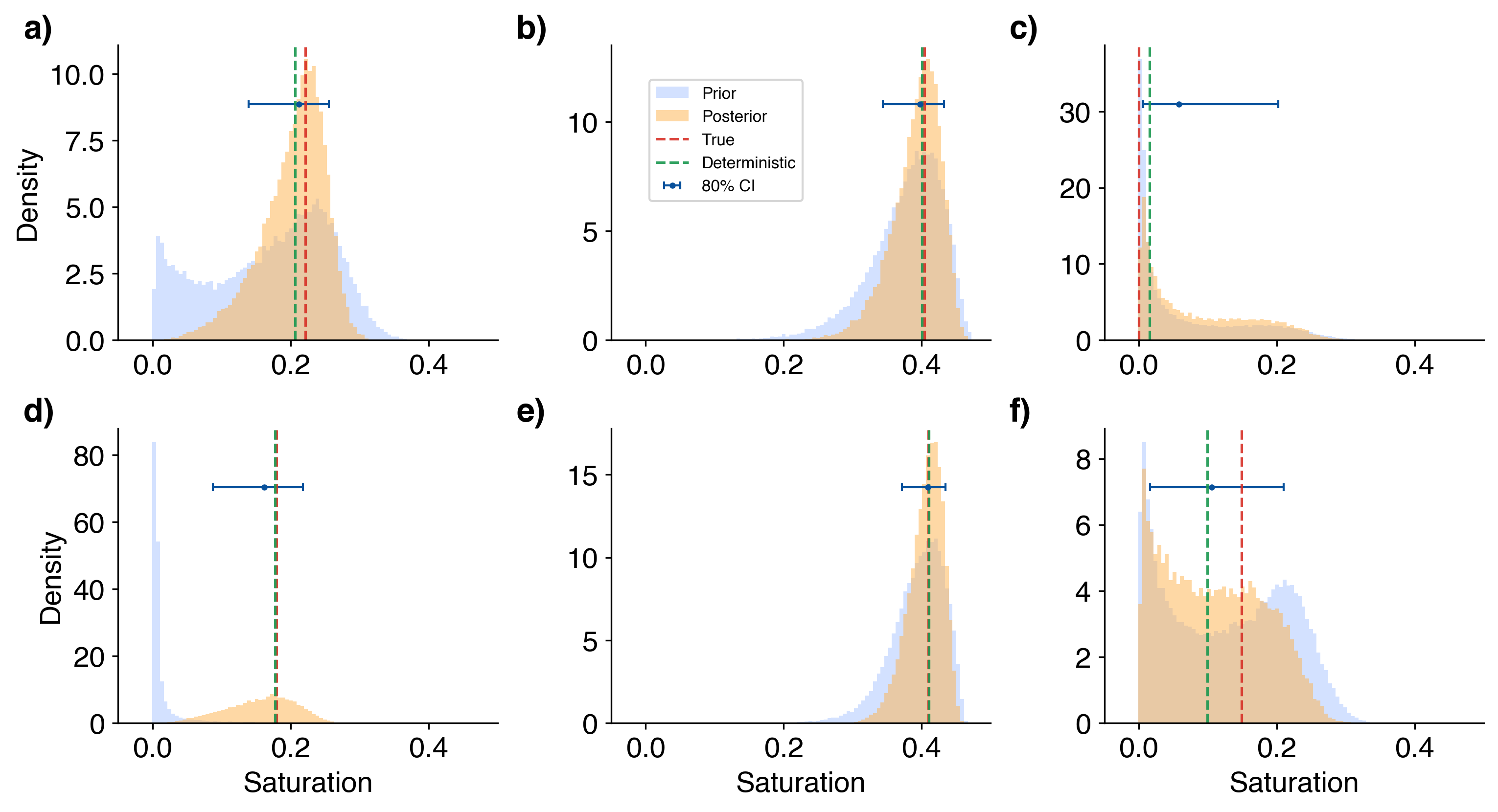}
\caption{
(a)-(f) Prior and posterior saturation distributions at six locations as labeled from 1 to 6 in Figure~\ref{fig:hmc_result}a.  
The red dotted line shows the true value, and the green dotted line shows the deterministic result. Horizontal bars denote the P$_{10}$-P$_{90}$ credible intervals.
}
\label{fig:hmc_hist}
\end{figure}

Because seismic illumination is spatially uneven, the inferred posterior exhibits corresponding variations in uncertainty and local behavior. Posterior marginal distributions at six locations marked in Figure~\ref{fig:hmc_result}a are further examined. As shown in Figure~\ref{fig:hmc_hist}b and Figure~\ref{fig:hmc_hist}e, near the injection well, the prior distributions are already narrow, and conditioning on the seismic data further reduces their variance. At the upper-left plume edge (Figure~\ref{fig:hmc_hist}a), the bimodal prior (with modes at 0 and 0.25) collapses into a unimodal posterior centered around 0.22, which is consistent with the true value (red dashed line). At the upper-right edge (Figure~\ref{fig:hmc_hist}c), both the prior and posterior remain close to zero, showing with high confidence that CO$_2$ did not reach this location. At the lower-left edge (Figure~\ref{fig:hmc_hist}d), the posterior shifts the prior from near-zero saturation toward approximately 0.18, correcting prior bias and agreeing with the true model. At the lower-right edge (Figure~\ref{fig:hmc_hist}f), both the prior and the inferred posterior remain broad (as indicated by the wide $P_{10}$-$P_{90}$ credible intervals), reflecting weak seismic sensitivity in this region. Additional data would be helpful to better constrain this portion of the model. 

The deterministic inversion generally agrees with the Bayesian results and corresponds roughly to the MAP estimate. It does, however, underestimate or overestimate saturation at certain locations (e.g., Figures~\ref{fig:hmc_result}a and \ref{fig:hmc_result}c). In contrast, the Bayesian results produce an entire distribution of plausible solutions rather than a single point estimate, from which uncertainty, confidence intervals, and parameter trade-offs can be directly assessed.

% Summary
In summary, the posterior distributions reliably recover CO$_2$ saturation changes, with low uncertainty near the injection well and higher uncertainty along plume margins where seismic constraints decrease. Incorporating seismic data provides spatial constraints that extend beyond those informed by geostatistical and fluid-flow priors. Because the framework is fully Bayesian, all uncertainty estimates arise from the explicit probabilistic combination of prior information and the seismic likelihood, rather than through heuristic tuning or regularization. As a result, the workflow delivers meaningful and interpretable uncertainty quantification through the joint assimilation of geostatistical, fluid-physics, and seismic information.

\section{Sensitivity and Uncertainty Analysis}
The previous sections demonstrated that the proposed Bayesian framework provides accurate plume monitoring under favorable conditions, including dense data coverage, noise-free observations, and well-matched prior assumptions. In realistic CO$_2$ storage operations, however, monitoring is challenged by sparse acquisition, variable data quality in time-lapse surveys, and unavoidable modeling errors. These factors introduce multiple sources of uncertainty that can affect the reliability of the inferred CO$_2$ plume. In this section, we assess several kinds of uncertainty. Some arise from incomplete knowledge or imperfect physical modeling, such as limitations in survey geometry and inaccuracies in rock physics or flow-simulation priors, generally referred to as epistemic uncertainty. Others stem from data noise caused by stochastic variations in field recordings (aleatoric uncertainty). Our goal is to identify the conditions under which the Bayesian framework remains robust and its quantified uncertainties provide added value over deterministic inversion. We further demonstrate that the workflow enables reliable plume monitoring, even with a single-source survey and in the presence of data noise, when the informative fluid-flow priors are incorporated.

\subsection{Probabilistic Assessment of Survey Geometry}
The monitoring results presented in Figure~\ref{fig:hmc_result} are obtained with five surface sources. In practice, however, long-term GCS monitoring often relies on sparse seismic acquisition due to cost constraints~\citep{yurikov2022seismic}. Such sparse data often render waveform monitoring highly ill-posed, making uncertainty quantification essential. Meanwhile, fewer sources substantially reduce the computational cost of Bayesian inversion, making probabilistic monitoring more affordable. Here, we use illustrative examples to show that Bayesian uncertainty quantification offers a meaningful framework for evaluating and thus guiding monitoring geometry design.

\begin{figure}
\centering
\includegraphics[width=0.93\textwidth]{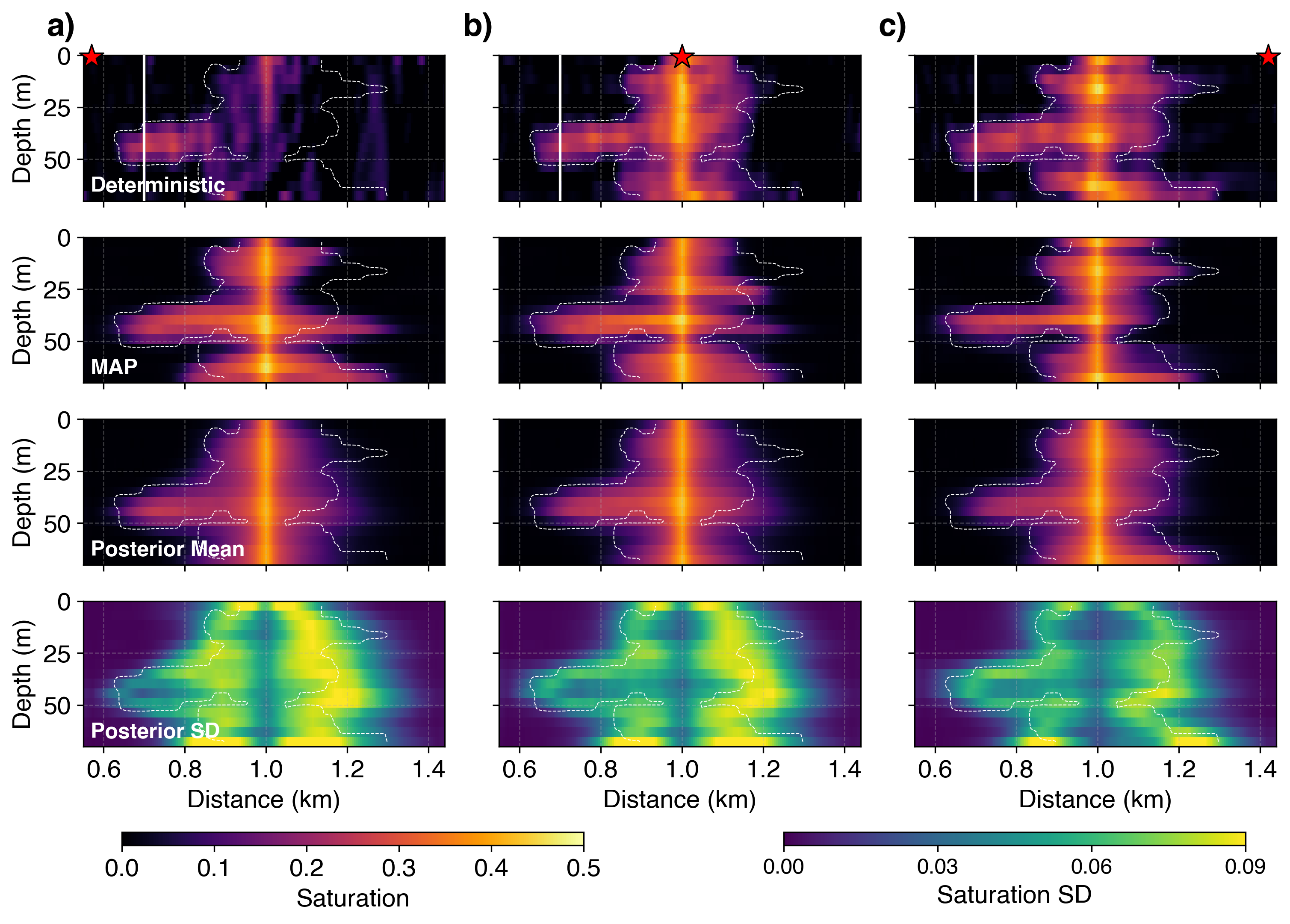}
\caption{Sensitivity of monitoring results to seismic survey geometry. Panels (a–c) show results for a single source placed at 0.5~km, 1.0~km, and 1.5~km, respectively. The sources (red stars) are deployed at the surface, with the left and right sources located outside the plotted domain. Each column presents, from top to bottom, the deterministic result, MAP estimate, posterior mean, and posterior standard deviation. The white vertical line denotes the monitoring well.
}
\label{fig:hmc_survey1}
\end{figure}

To investigate the impact of survey geometry, we analyze the extreme single-source scenarios using the same model shown in Figure~\ref{fig:model}. Each scenario employs only one source placed at 0.5~km, 1.0~km, or 1.5~km (Figures~\ref{fig:hmc_survey1}a-c), with a 30~Hz Ricker wavelet and the same monitoring well. The deterministic result fails to recover the plume when illumination is extremely poor (Figure~\ref{fig:hmc_survey1}a), whereas the results improve as the source moves to 1.0~km and 1.5~km (Figures~\ref{fig:hmc_survey1}b-c). In contrast, the Bayesian monitoring approach yields interpretable plume structures across all three cases, as shown by the MAP estimates and posterior means in Figure~\ref{fig:hmc_survey1}. As reflected in the posterior standard deviations, the source at 0.5~km produces poor illumination and consequently large posterior variance along both plume wings. The 1.0~km source better resolves the left plume wing but leaves substantial uncertainty on the right. The source at 1.5~km performs best among the three cases due to better illumination, recovering both plume wings and producing the lowest uncertainties. It is the most effective source location among the three tested surveys.

\begin{figure}
\centering
\includegraphics[width=0.93\textwidth]{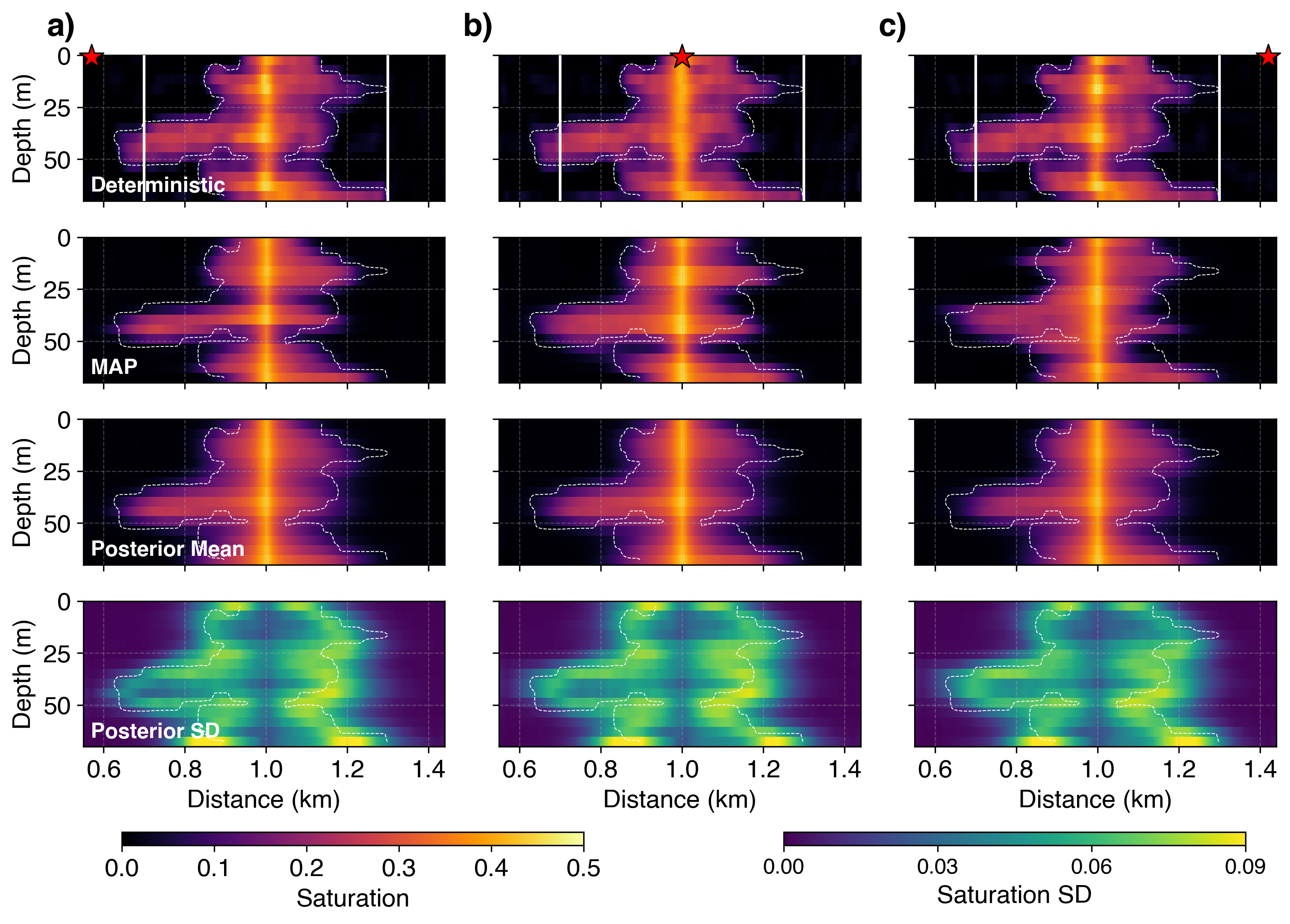}
\caption{Same as Figure~\ref{fig:hmc_survey1}, but with two monitoring wells positioned at 0.7 km and 1.3 km.}
\label{fig:hmc_survey2}
\end{figure}

We also evaluated how adding a second monitoring well could improve performance. Although drilling new wells is costly, many storage sites already have multiple existing wells, making this comparison relevant in practice. In our test, a second well is placed at 1.3~km with the same depth range as the first. As shown in Figure~\ref{fig:hmc_survey2}, deterministic results improve due to the increased data coverage. Bayesian inference similarly benefits: posterior uncertainties drop substantially for the 0.5~km and 1.0~km source cases. To interpret these trends, recall that additional measurements contribute new Fisher information, which quantifies how strongly the data constrain the model parameters. When the survey geometry already provides sufficient constraints, as in the 1.5~km source case, the Fisher information is nearly maximized, so adding another monitoring well yields only marginal improvement. This probabilistic perspective suggests that, once an effective source location is selected, drilling an additional well offers limited value because it contributes little additional Fisher information. Judicious survey design can therefore provide greater gains relative to its cost. A comprehensive optimization, however, would also need to incorporate site logistics, operational constraints, and risk considerations, factors beyond the scope of the illustrative examples presented here.

\subsection{Sensitivity to Data Noise}
We now examine how measurement noise further affects detectability in the extremely sparse, single-source monitoring scenario. We use the best single-source case as identified above (1.5~km source case). In this analysis, realistic noise extracted from land seismic recordings (Supporting Figure~S6) is added to both baseline and monitor data at signal-to-noise ratios (SNRs) of 30 dB and 25 dB. These levels approximate field conditions in which repeatability is limited and weak time-lapse signals are easily masked.

\begin{figure}[H]
\centering
\includegraphics[width=0.93\textwidth]{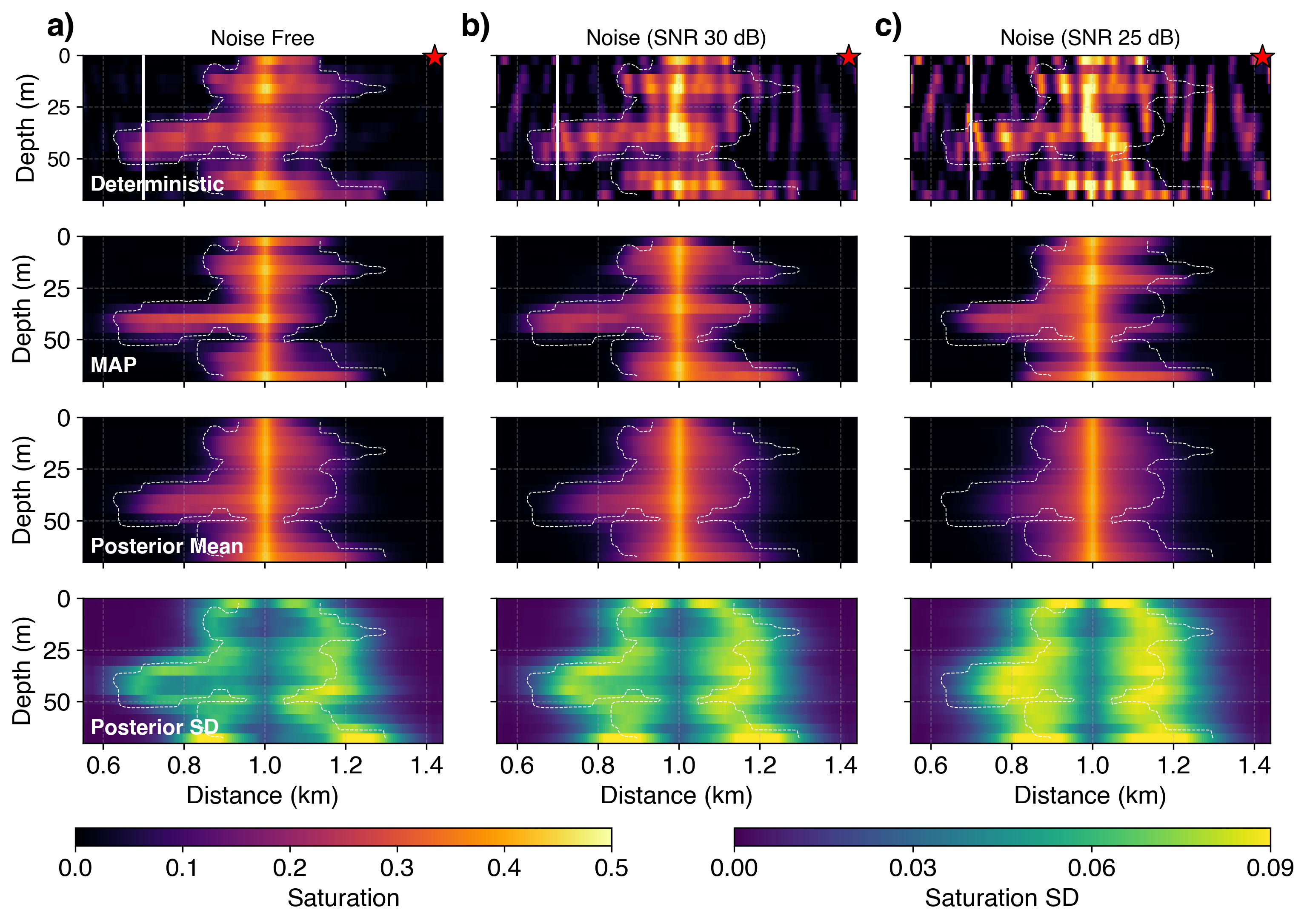}
\caption{Sensitivity of monitoring results to measurement noise. Comparison of inversion results for a single source under three noise conditions: noise-free (a), 30~dB noise (b), and 25~dB noise (c). The dashed line denotes the true CO$_2$ plume boundary. }
\label{fig:hmc_noise}
\end{figure}

\begin{figure}
\centering
\includegraphics[width=0.7\textwidth]{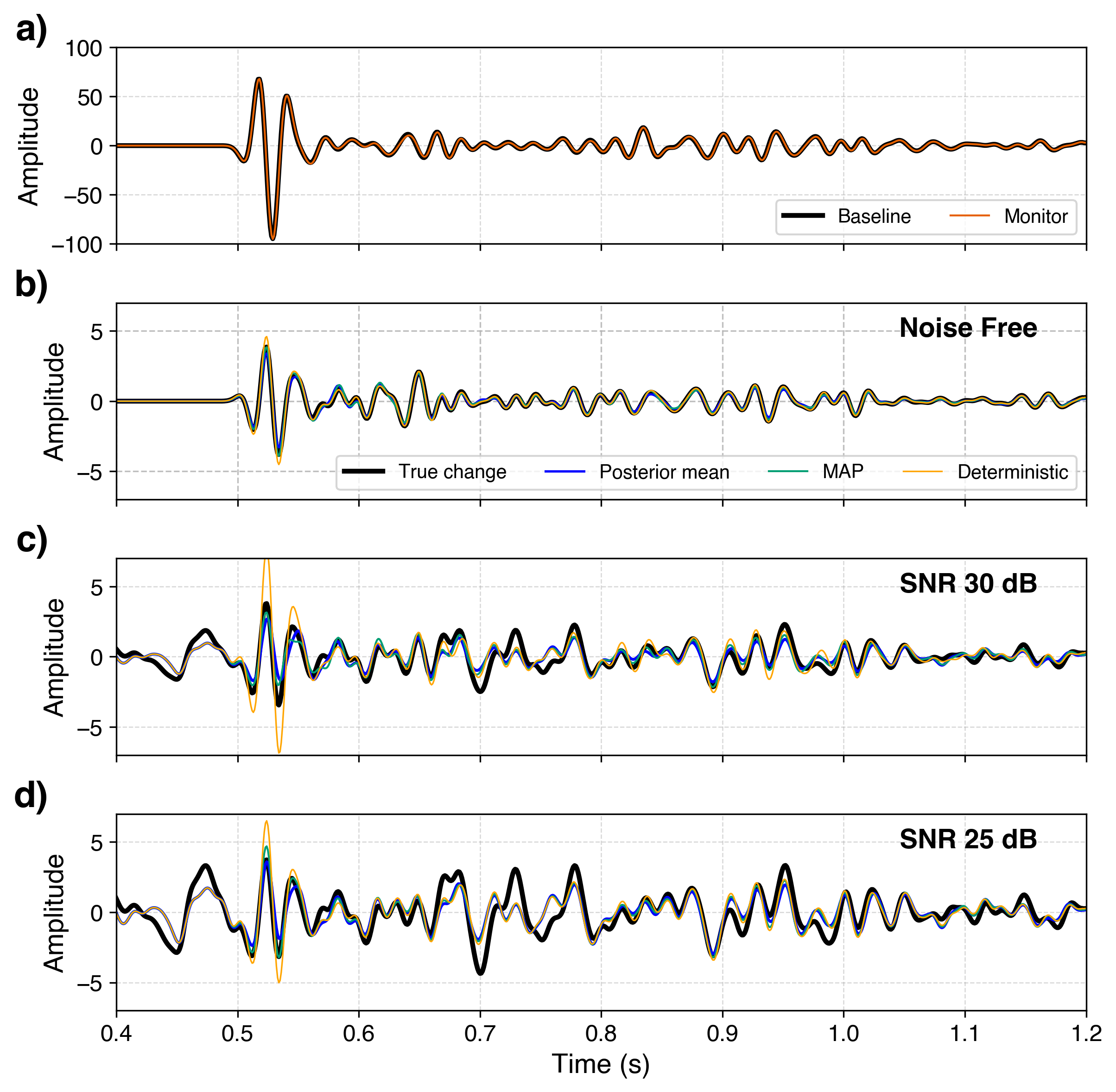}
\caption{Posterior waveform predictions at a receiver located at a depth of 1.4~km. 
(a) Baseline and monitor waveforms for reference (without noise).
(b) True time-lapse difference (monitor minus baseline) for the noise-free case, compared with the posterior predictions, which include the posterior mean, MAP estimate, and deterministic result.
(c-d) Same comparison as in panel (b), but for data contaminated with 30 dB and 25 dB noise, respectively. Note that the amplitudes of waveform changes are amplified.
}
\label{fig:hmc_waveform}
\end{figure}

As shown in Figure~\ref{fig:hmc_noise}, introducing noise at different levels renders the deterministic monitoring results unstable and causes them to be dominated by artifacts that could be misinterpreted as leakage. Bayesian inference, however, remains substantially more robust. Both the MAP estimate and posterior mean still retrieve the main CO$_2$ plume body, although the posterior variance increases, particularly along the plume boundaries. Boundary regions usually exhibit smaller saturation contrasts, produce weaker scattered wavefields, and consequently have lower SNR. The posterior ensemble reflects this reduced Fisher information by showing higher uncertainty where data sensitivity is low. Waveform comparisons are shown in Figure \ref{fig:hmc_waveform}. In the noise-free case, the true time-lapse difference (Figure \ref{fig:hmc_waveform}b), though small relative to the absolute waveform (Figure \ref{fig:hmc_waveform}a), is accurately predicted by both deterministic and Bayesian predictions. With 30 dB and 25 dB noise, the observed differences become heavily contaminated: deterministic inversion becomes unstable, whereas Bayesian predictions still recover the coherent time-lapse signal (Figure \ref{fig:hmc_waveform}c–d) due to the informative prior.

These results show that measurement noise impacts uncertainty in regions of low seismic sensitivity. Although single-source monitoring with realistic noise is generally inadequate for conventional deterministic monitoring, our Bayesian framework remains robust and provides interpretable results with quantified uncertainty. This offers guidance for assessing detectability and repeatability in time-lapse seismic monitoring of CO$_2$ storage sites.

\subsection{Sensitivity to Modeling Errors}
Beyond data-related uncertainties, CO$_2$ monitoring is also affected by systematic modeling errors arising from discrepancies between the physical processes assumed in the inversion workflow and those governing the true subsurface. Such modeling errors can stem from uncertainty in fluid-flow physics and imperfect rock physics relationships. To assess the robustness of the proposed Bayesian framework under these conditions, we evaluate two representative forms of modeling error: (1) prior-target mismatch in plume size, and (2) biased rock physics mapping.

\begin{figure}[H]
\centering
\includegraphics[width=0.93\textwidth]{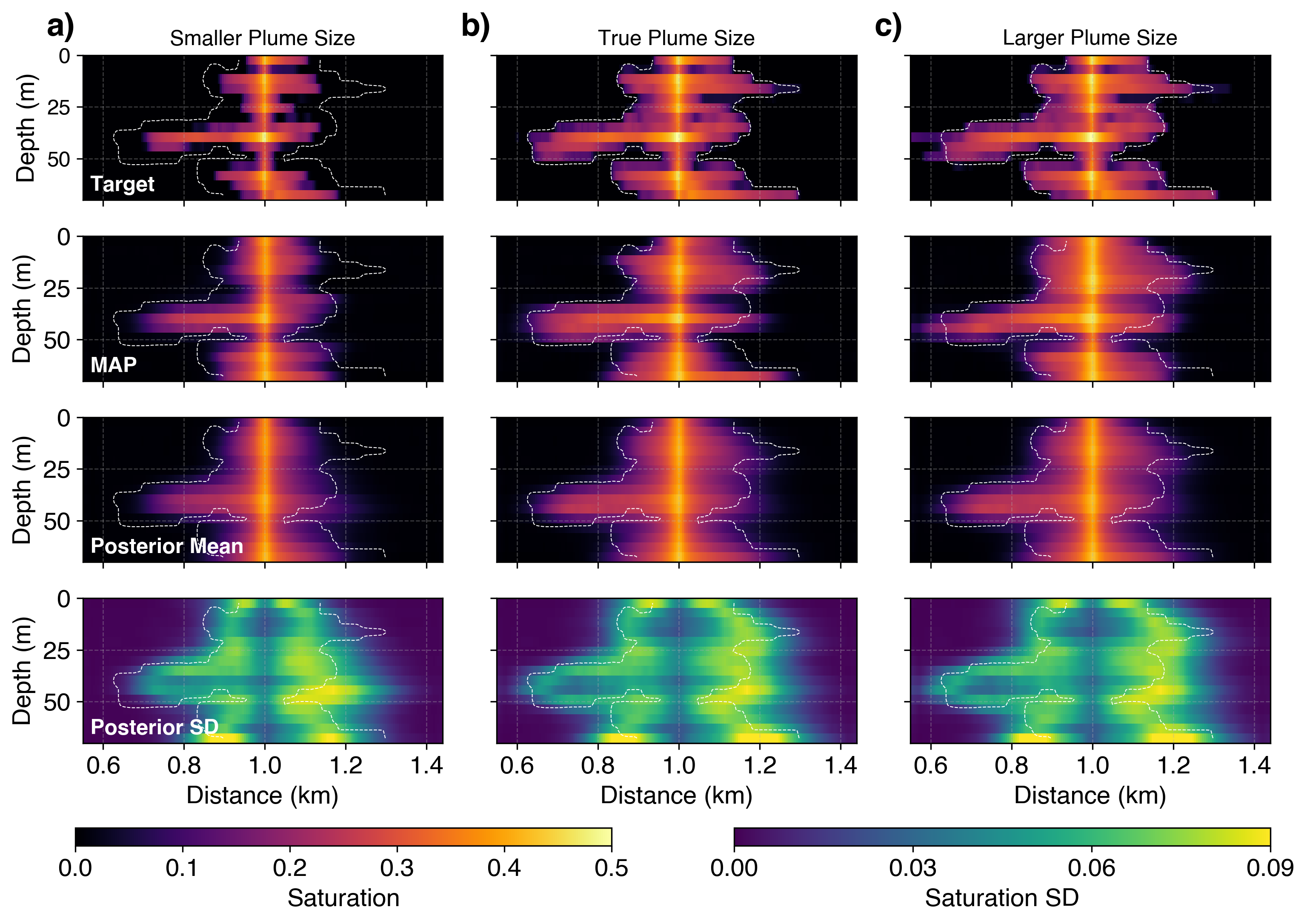}
\caption{Sensitivity to prior-target mismatch. (a-c) True saturation fields at 4, 10 (training snapshot), and 12 months of injection. Rows show the MAP estimate, posterior mean, and posterior standard deviation for each case.}
\label{fig:hmc_prior_error}
\end{figure}

The VAE prior in this study was trained on saturation field snapshots from 10 months of injection. In practice, plume size evolves over time, and the true saturation field may fall partially or even entirely outside the support of the learned prior. To evaluate this effect, we apply the monitoring workflow to a smaller plume (4 months) and a larger plume (12 months) while still using the 10-month prior. As shown in Figure~\ref{fig:hmc_prior_error}, the Bayesian inversion reliably recovers the plume structure at both earlier and moderately later stages. This robustness arises from the variability of plume patterns encoded in the prior, which already spans a range of plume sizes around the 10-month snapshot. When the plume becomes far smaller or larger than those represented in the prior, however, posterior samples begin to miss the full lateral extent. A broader prior, trained across multiple injection stages or incorporating variability in key physical parameters, would help mitigate this modeling error. However, broader priors are less informative. There is always an inherent trade-off between prior expressiveness and inversion performance.

\begin{figure}
\centering
\includegraphics[width=0.8\textwidth]{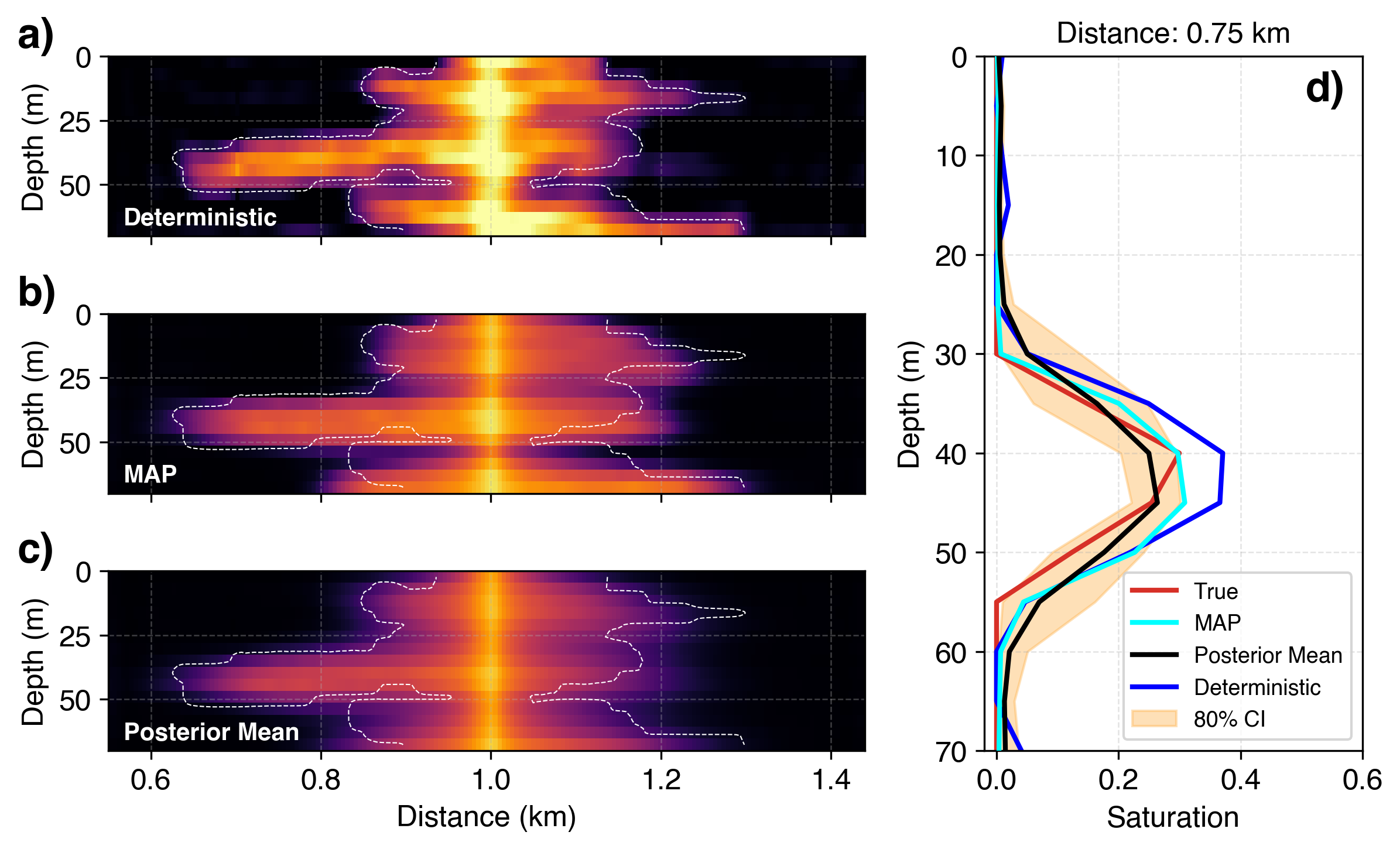}
\caption{Sensitivity to rock physics bias, where observed data use a Brie's coefficient of 3 while monitoring uses 2.
(a) Deterministic inversion with incorrect rock physics mapping.
(b) MAP estimate.
(c) Posterior mean.
(d) Vertical saturation profile at 0.75 km comparing all methods with the true value.}
\label{fig:hmc_rock_error}
\end{figure}

We next evaluate the sensitivity of the Bayesian framework to errors in the rock physics model, which links CO$_2$ saturation to seismic velocities. Any bias in this mapping introduces a form of model error that propagates through the likelihood and affects the posterior. To introduce a controlled bias, the “true” seismic data are generated using a Brie's coefficient of 3 in Eq.~\ref{eq:brie}, whereas the Bayesian inference assumes an coefficient of 2, creating a systematic rock physics mismatch. Figure~\ref{fig:hmc_rock_error} summarizes the results. Deterministic inversion systematically overestimates CO$_2$ saturation because the incorrect mapping converts the observed velocity reduction into high saturation values. In contrast, the Bayesian MAP and posterior mean remain closer to the true plume, and the posterior $P_{10}$–$P_{90}$ interval consistently brackets the correct saturation profile (Figure~\ref{fig:hmc_rock_error}d). This robustness stems from the variability encoded in the generative prior, whose broad representation of plausible plume shapes helps mitigate errors introduced by an imperfect rock physics model.

These tests show that Bayesian monitoring with a fluid-flow prior is tolerant of moderate modeling errors, including plume-size mismatch and rock physics uncertainty. When rock physics errors are present, deterministic inversion fails because the mapping is fixed and cannot adapt to any bias, whereas the Bayesian approach retains some flexibility. In principle, this framework could be extended to jointly infer uncertain rock physics parameters, such as the Brie's coefficient, together with the saturation field in a fully probabilistic manner~\citep{li2020coupled}.

\section{Discussion}

\subsection{Prior Construction and Representation}
% Realistic geo-model priors to fit a specific field data
The effectiveness of Bayesian seismic monitoring depends on constructing realistic priors. For a particular site, geological interpretations, including depositional settings, lithological classifications, and facies distributions, provide essential inputs for constructing the prior. Geostatistical simulation approaches, such as object-based simulation~\citep{wang2018conditioning} and sequential Gaussian simulation~\citep{deutsch1992geostatistical, manchuk2012flexible}, can then be utilized to generate plausible geological models. Additionally, some local measurements (such as well logging data and rock core samples) could also be used as conditioning data to constrain stochastic realizations. With appropriate prior geomodels, the reservoir simulation results can characterize potential flow behaviors. 

% stochastic rock physics modeling
Another component of prior design is rock physics modeling, which links fluid-flow simulations to seismic velocity. In this study, we employed a deterministic rock physics model to transform generated CO$_2$ saturation fields into P-wave velocity changes. Incorporating stochastic rock physics modeling could allow these variations to be considered explicitly, help reduce the risk of biasing the inversion toward a single petrophysical scenario. In addition, the seismically inverted saturation fields could be incorporated into history matching or data assimilation workflows to calibrate geomodels and further reduce geological uncertainty~\citep{wang2025deep}. The history-matched posterior flow simulations using the observed saturation and pressure data~\citep[e.g.,][]{han2024surrogate} could then serve as prior ensembles for subsequent seismic monitoring, progressively refining CO$_2$ saturation estimates. In this way, geological, petrophysical, and seismic uncertainties can be integrated into a closed-loop assimilation framework for assessments of CO$_2$ storage performance.

% prior representation
In terms of prior representation, using a generative and compact parameterization improves sampling efficiency in high-dimensional inverse problems. In this study, we applied a VAE to compress CO$_2$ saturation realizations into a low-dimensional latent space, effectively mapping the original model domain into a Gaussian-distributed latent space (see Supporting Figure~S4). Alternative generative strategies, such as diffusion-based models~\citep{ho2020denoising, wang2024controllable, xu2024diffsim, di2025latent}, offer even higher fidelity and can be capable of capturing complex spatial correlations and multi-modal distributions. Nevertheless, in practice, it remains essential to balance generation quality and computational speed when selecting an appropriate generative prior strategy for computationally intensive Bayesian inference.

\subsection{Bayesian Inference Strategies}
Bayesian inference offers a way to quantify uncertainty in seismic monitoring, but the choice of sampling strategy determines the balance between accuracy and computational cost. HMC is attractive because it leverages Hamiltonian dynamics to avoid the random-walk behavior of classic MCMC, thus achieving higher acceptance rates (up to 65\% compared to 23\% for standard Metropolis-Hastings) and scaling more favorably with dimensionality~\citep{neal2011mcmc}. Still, while HMC is asymptotically exact for high-dimensional problems, its performance depends on careful tuning of hyperparameters such as the step size, trajectory length, and mass matrix. Several extensions have been developed to address these challenges, including the No-U-Turn Sampler, which automatically selects trajectory lengths and reduces the need for manual tuning~\citep{hoffman2014no}. Autotuning strategies for updating the mass matrix~\citep{fichtner2021autotuning} can further improve sampling performance. In addition, tempered-posterior formulations and multi-chain sampling can help identify separated modes in complex landscape. These advanced strategies make HMC increasingly practical for large-scale and automated Bayesian monitoring, and integrating them into our workflow is promising for future work.

Other Bayesian inference methods trade some accuracy for computational speed. Variational inference (VI) approximates the posterior with a parameterized distribution family and can converge orders of magnitude faster than HMC~\citep{blei2017variational}, though it is not asymptotically exact and may struggle with multi-modal or heavy-tailed posteriors. Stein variational gradient descent (SVGD)~\citep{liu2016stein} provides a particle-based strategy, transporting particles toward the posterior through deterministic updates. Although SVGD is more efficient than traditional MCMC, its performance is sensitive to kernel and bandwidth choices, especially in high-dimensional model spaces.

The appropriate Bayesian inference strategy depends on the complexity of the monitoring target and the available computational resources. When using a VAE model representation that maps highly correlated physical parameters into an approximately Gaussian latent space, methods such as VI or SVGD may become particularly appealing. Although in this study we adopted the asymptotically exact HMC sampler, future work should benchmark alternative approaches to identify the optimal balance between computational efficiency and statistical reliability.

\subsection{Other Future Efforts}
This study is based on 2-D acoustic monitoring, where P-wave velocity is treated as the quantity of interest linked to CO$_2$ saturation. In reality, CO$_2$ plume migration influences not only P-wave velocity but also S-wave velocity, density, and attenuation, with these properties coupled through fluid substitution and pore-pressure variations. Elastic multi-parameter and fully 3-D monitoring would provide better constraints on fluid-related and stress-induced changes, thus better supporting subsequent petrophysical and geomechanical interpretations. Meanwhile, the strong trade-offs between multiple parameters make the problem more challenging. Posterior analysis can be useful in such cases to identify parameter trade-offs and can thus help distinguish different subsurface processes.
 
Finally, applying the proposed framework to field data will be an important direction for future work. Field datasets introduce additional complexities, such as inaccuracies in the baseline model, variations in source signatures, and near-surface conditions. Addressing these challenges requires not only a robust seismic inference framework but also improved data preprocessing and denoising strategies.

\section{Conclusions}
We presented a Bayesian seismic monitoring framework that incorporates geostatistically generated geomodels, fluid-flow simulation, rock physics modeling, and generative deep learning to improve time-lapse CO$_2$ plume characterization. The reservoir simulation ensemble provides physics-based priors, and the VAE compresses their high-dimensional variability into a low-dimensional latent space with an approximately Gaussian distribution. This representation enables the posterior to be explored with the HMC sampler at manageable cost (this cost would be prohibitive in the original model space). The numerical experiments show that conditioning informative priors on seismic data enables reliable recovery of plume geometry, while the resulting posterior variance reflects spatial differences in seismic sensitivity. Importantly, the developed method remains stable under extremely sparse acquisition and realistic noise, while the deterministic approach may become unreliable in such cases. The quantified posterior uncertainties can be used to guide survey design by revealing where additional measurements would most effectively reduce uncertainty. In addition, the statistical monitoring can act to mitigate biases arising from imperfect rock physics modeling. Overall, coupling reservoir physics with generative latent-space priors yields a robust and flexible Bayesian seismic monitoring framework. The methodology can be extended to elastic and 3-D formulations and enhanced to integrate additional data.

%%%%%%%%%%%%%%%%%%%%%%%%%%%%%%%%%%%%%%%%%%%%%%%
%
% DATA SECTION and ACKNOWLEDGMENTS
%
%%%%%%%%%%%%%%%%%%%%%%%%%%%%%%%%%%%%%%%%%%%%%%%

\section*{Open Research Section}
The codebase, documentation, and simulated CO$_2$ plume required to reproduce all figures and results presented in this manuscript are already archived at the persistent repository ({https://doi.org/10.5281/zenodo.17917000}) \citep{li_2025_17917000} and will be made publicly available upon potential acceptance. 

\acknowledgments
We thank the Stanford Earth imaging Project (SEP) affiliate companies and the Stanford Center for Carbon Storage (SCCS) affiliate companies for their financial support. We also thank Oleg Volkov for assistance with GEOS and Robert Clapp for discussions on HMC. The authors acknowledge the use of AI tools to assist with language refinement.

\bibliography{ref}

@article{aragao2020elastic,
  title={Elastic full-waveform inversion with probabilistic petrophysical model constraints},
  author={Aragao, Odette and Sava, Paul},
  journal={Geophysics},
  volume={85},
  number={2},
  pages={R101--R111},
  year={2020},
  publisher={Society of Exploration Geophysicists}
}

@article{jenkins2020state,
  title={The State of the Art in Monitoring and Verification: an update five years on},
  author={Jenkins, Charles},
  journal={International Journal of Greenhouse Gas Control},
  volume={100},
  pages={103118},
  year={2020},
  publisher={Elsevier}
}

@book{davis2019geophysics,
  title={Geophysics and Geosequestration},
  author={Davis, Thomas L and Landr{\o}, Martin and Wilson, Malcolm},
  year={2019},
  publisher={Cambridge University Press}
}

@article{chadwick2009latest,
  title={Latest time-lapse seismic data from {S}leipner yield new insights into {CO$_2$} plume development},
  author={Chadwick, RA and Noy, D and Arts, R and Eiken, O},
  journal={Energy Procedia},
  volume={1},
  number={1},
  pages={2103--2110},
  year={2009},
  publisher={Elsevier}
}

@article{hicks2016time,
  title={Time-lapse full-waveform inversion as a reservoir-monitoring tool—A North Sea case study},
  author={Hicks, Erik and Hoeber, Henning and Houbiers, Marianne and Lescoffit, S{\'e}verine Pannetier and Ratcliffe, Andrew and Vinje, Vetle},
  journal={The Leading Edge},
  volume={35},
  number={10},
  pages={850--858},
  year={2016},
  publisher={Society of Exploration Geophysicists}
}

@article{roach2018evolution,
  title={Evolution of a deep {CO$_2$} plume from time-lapse seismic imaging at the Aquistore storage site, Saskatchewan, {Canada}},
  author={Roach, Lisa AN and White, DJ},
  journal={International Journal of Greenhouse Gas Control},
  volume={74},
  pages={79--86},
  year={2018},
  publisher={Elsevier}
}

@article{pevzner2021seismic,
  title={Seismic monitoring of a small {CO$_2$} injection using a multi-well DAS array: Operations and initial results of Stage 3 of the {CO2CRC Otway} project},
  author={Pevzner, Roman and Isaenkov, Roman and Yavuz, Sinem and Yurikov, Alexey and Tertyshnikov, Konstantin and Shashkin, Pavel and Gurevich, Boris and Correa, Julia and Glubokovskikh, Stanislav and Wood, Todd and others},
  journal={International Journal of Greenhouse Gas Control},
  volume={110},
  pages={103437},
  year={2021},
  publisher={Elsevier}
}

@article{asnaashari2015time,
  title={Time-lapse seismic imaging using regularized full-waveform inversion with a prior model: which strategy?},
  author={Asnaashari, Amir and Brossier, Romain and Garambois, St{\'e}phane and Audebert, Fran{\c{c}}ois and Thore, Pierre and Virieux, Jean},
  journal={Geophysical Prospecting},
  volume={63},
  number={1},
  pages={78--98},
  year={2015},
  publisher={European Association of Geoscientists \& Engineers}
}

@article{egorov2017time,
  title={Time-lapse full waveform inversion of vertical seismic profile data: workflow and application to the {CO2CRC Otway} project},
  author={Egorov, Anton and Pevzner, Roman and B{\'o}na, Andrej and Glubokovskikh, Stanislav and Puzyrev, Vladimir and Tertyshnikov, Konstantin and Gurevich, Boris},
  journal={Geophysical Research Letters},
  volume={44},
  number={14},
  pages={7211--7218},
  year={2017},
  publisher={Wiley Online Library}
}

@book{tarantola2005inverse,
  title={Inverse problem theory and methods for model parameter estimation},
  author={Tarantola, Albert},
  year={2005},
  publisher={Society for Industrial and Applied Mathematics}
}

@article{sambridge2002monte,
  title={{M}onte {C}arlo methods in geophysical inverse problems},
  author={Sambridge, Malcolm and Mosegaard, Klaus},
  journal={Reviews of Geophysics},
  volume={40},
  number={3},
  pages={3--1},
  year={2002},
  publisher={Wiley Online Library}
}

@article{isaenkov2021automated,
  title={An automated system for continuous monitoring of {CO$_2$} geosequestration using multi-well offset VSP with permanent seismic sources and receivers: Stage 3 of the {CO2CRC Otway} Project},
  author={Isaenkov, Roman and Pevzner, Roman and Glubokovskikh, Stanislav and Yavuz, Sinem and Yurikov, Alexey and Tertyshnikov, Konstantin and Gurevich, Boris and Correa, Julia and Wood, Todd and Freifeld, Barry and others},
  journal={International Journal of Greenhouse Gas Control},
  volume={108},
  pages={103317},
  year={2021},
  publisher={Elsevier}
}

@article{maharramov2016time,
  title={Time-lapse inverse theory with applications},
  author={Maharramov, Musa and Biondi, Biondo L and Meadows, Mark A},
  journal={Geophysics},
  volume={81},
  number={6},
  pages={R485--R501},
  year={2016},
  publisher={Society of Exploration Geophysicists}
}

@article{batzle1992seismic,
  title={Seismic properties of pore fluids},
  author={Batzle, Michael and Wang, Zhijing},
  journal={Geophysics},
  volume={57},
  number={11},
  pages={1396--1408},
  year={1992},
  publisher={Society of Exploration Geophysicists}
}

@article{gebraad2020bayesian,
  title={Bayesian elastic full-waveform inversion using {H}amiltonian {Monte Carlo}},
  author={Gebraad, Lars and Boehm, Christian and Fichtner, Andreas},
  journal={Journal of Geophysical Research: Solid Earth},
  volume={125},
  number={3},
  pages={e2019JB018428},
  year={2020},
  publisher={Wiley Online Library}
}

@article{zunino2023hmclab,
  title={{HMCLab}: a framework for solving diverse geophysical inverse problems using the {H}amiltonian {Monte Carlo} method},
  author={Zunino, Andrea and Gebraad, Lars and Ghirotto, Alessandro and Fichtner, Andreas},
  journal={Geophysical Journal International},
  volume={235},
  number={3},
  pages={2979--2991},
  year={2023},
  publisher={Oxford University Press}
}

@article{neal2011mcmc,
  title={{MCMC} using {H}amiltonian dynamics},
  author={Neal, Radford M and others},
  journal={Handbook of {M}arkov chain {Monte Carlo}},
  volume={2},
  number={11},
  pages={2},
  year={2011},
  publisher={Chapman and Hall/CRC}
}

@article{zhao2021gradient,
  title={A gradient-based {M}arkov chain {Monte Carlo} method for full-waveform inversion and uncertainty analysis},
  author={Zhao, Zeyu and Sen, Mrinal K},
  journal={Geophysics},
  volume={86},
  number={1},
  pages={R15--R30},
  year={2021},
  publisher={Society of Exploration Geophysicists}
}

@article{zhang2020seismic,
  title={Seismic tomography using variational inference methods},
  author={Zhang, Xin and Curtis, Andrew},
  journal={Journal of Geophysical Research: Solid Earth},
  volume={125},
  number={4},
  pages={e2019JB018589},
  year={2020},
  publisher={Wiley Online Library}
}

@article{metropolis1953equation,
  title={Equation of state calculations by fast computing machines},
  author={Metropolis, Nicholas and Rosenbluth, Arianna W and Rosenbluth, Marshall N and Teller, Augusta H and Teller, Edward},
  journal={The Journal of Chemical Physics},
  volume={21},
  number={6},
  pages={1087--1092},
  year={1953},
  publisher={American Institute of Physics}
}

@article{hoffman2014no,
  title={The {No-U-Turn} sampler: adaptively setting path lengths in {H}amiltonian {Monte Carlo}},
  author={Hoffman, Matthew D and Gelman, Andrew and others},
  journal={Journal of Machine Learning Research},
  volume={15},
  number={1},
  pages={1593--1623},
  year={2014}
}

@article{fichtner2021autotuning,
  title={Autotuning {H}amiltonian {Monte Carlo} for efficient generalized nullspace exploration},
  author={Fichtner, Andreas and Zunino, Andrea and Gebraad, Lars and Boehm, Christian},
  journal={Geophysical Journal International},
  volume={227},
  number={2},
  pages={941--968},
  year={2021},
  publisher={Oxford University Press}
}

@article{zhang2018multiparameter,
  title={Multiparameter elastic full waveform inversion with facies-based constraints},
  author={Zhang, Zhendong and Alkhalifah, Tariq and Naeini, Ehsan Zabihi and Sun, Bingbing},
  journal={Geophysical Journal International},
  volume={213},
  number={3},
  pages={2112--2127},
  year={2018},
  publisher={Oxford University Press}
}

@article{li2016integrated,
  title={Integrated VTI model building with seismic data, geologic information, and rock-physics modeling—Part 1: Theory and synthetic test},
  author={Li, Yunyue and Biondi, Biondo and Clapp, Robert and Nichols, Dave},
  journal={Geophysics},
  volume={81},
  number={5},
  pages={C177--C191},
  year={2016},
  publisher={Society of Exploration Geophysicists}
}

@article{zhu2016bayesian,
  title={A {Bayesian} approach to estimate uncertainty for full-waveform inversion using a priori information from depth migration},
  author={Zhu, Hejun and Li, Siwei and Fomel, Sergey and Stadler, Georg and Ghattas, Omar},
  journal={Geophysics},
  volume={81},
  number={5},
  pages={R307--R323},
  year={2016},
  publisher={Society of Exploration Geophysicists}
}

@article{span1996new,
  title={A new equation of state for carbon dioxide covering the fluid region from the triple-point temperature to 1100 {K} at pressures up to 800 {MPa}},
  author={Span, Roland and Wagner, Wolfgang},
  journal={Journal of Physical and Chemical Reference Data},
  volume={25},
  number={6},
  pages={1509--1596},
  year={1996},
  publisher={American Institute of Physics for the National Institute of Standards and~…}
}

@article{kingma2013auto,
  title={Auto-encoding Variational {Bayes}},
  author={Kingma, Diederik P and Welling, Max},
  journal={arXiv preprint arXiv:1312.6114},
  year={2013}
}

@article{wang2025deep,
  title={Deep learning framework for history matching {CO$_2$} storage with {4D} seismic and monitoring well data},
  author={Wang, Nanzhe and Durlofsky, Louis J},
  journal={Geoenergy Science and Engineering},
  volume={248},
  pages={213736},
  year={2025},
  publisher={Elsevier}
}

@book{remy2009applied,
  title={Applied geostatistics with {SGeMS}: A user's guide},
  author={Remy, Nicolas and Boucher, Alexandre and Wu, Jianbing},
  year={2009},
  publisher={Cambridge University Press}
}

@article{liu2016stein,
  title={Stein variational gradient descent: A general purpose {Bayesian} inference algorithm},
  author={Liu, Qiang and Wang, Dilin},
  journal={Advances in Neural Information Processing Systems},
  volume={29},
  year={2016}
}

@article{blei2017variational,
  title={Variational inference: A review for statisticians},
  author={Blei, David M and Kucukelbir, Alp and McAuliffe, Jon D},
  journal={Journal of the American statistical Association},
  volume={112},
  number={518},
  pages={859--877},
  year={2017},
  publisher={Taylor \& Francis}
}

@article{yurikov2022seismic,
  title={Seismic monitoring of {CO$_2$} geosequestration using multi-well {4D DAS VSP}: Stage 3 of the {CO2CRC Otway} project},
  author={Yurikov, Alexey and Tertyshnikov, Konstantin and Yavuz, Sinem and Shashkin, Pavel and Isaenkov, Roman and Sidenko, Evgenii and Glubokovskikh, Stanislav and Barraclough, Paul and Pevzner, Roman},
  journal={International Journal of Greenhouse Gas Control},
  volume={119},
  pages={103726},
  year={2022},
  publisher={Elsevier}
}

@article{ho2020denoising,
  title={Denoising diffusion probabilistic models},
  author={Ho, Jonathan and Jain, Ajay and Abbeel, Pieter},
  journal={Advances in Neural Information Processing Systems},
  volume={33},
  pages={6840--6851},
  year={2020}
}

@article{wang2024controllable,
  title={Controllable seismic velocity synthesis using generative diffusion models},
  author={Wang, Fu and Huang, Xinquan and Alkhalifah, Tariq},
  journal={Journal of Geophysical Research: Machine Learning and Computation},
  volume={1},
  number={3},
  pages={e2024JH000153},
  year={2024},
  publisher={Wiley Online Library}
}

@article{queisser2013full,
  title={Full waveform inversion in the time lapse mode applied to {CO$_2$} storage at {S}leipner},
  author={Quei{\ss}er, Manuel and Singh, Satish C},
  journal={Geophysical Prospecting},
  volume={61},
  number={3},
  pages={537--555},
  year={2013},
  publisher={European Association of Geoscientists \& Engineers}
}

@article{kirkpatrick1983optimization,
  title={Optimization by simulated annealing},
  author={Kirkpatrick, Scott and Gelatt Jr, C Daniel and Vecchi, Mario P},
  journal={Science},
  volume={220},
  number={4598},
  pages={671--680},
  year={1983},
  publisher={American Association for the Advancement of Science}
}

@inproceedings{li2024time,
  title={Time-lapse full-waveform inversion by model order reduction using radial basis function},
  author={Li, Haipeng and Clapp, Robert G},
  booktitle={Fourth {I}nternational {M}eeting for {A}pplied {G}eoscience \& {E}nergy},
  pages={807--811},
  year={2024},
  organization={Society of Exploration Geophysicists and American Association of Petroleum}
}

@article{settgast2024geos,
  title={{GEOS}: A performance portable multi-physics simulation framework for subsurface applications},
  author={Settgast, Randolph R and Aronson, Ryan M and Besset, Julien R and Borio, Andrea and Bui, Quan M and Byer, Thomas J and Castelletto, Nicola and Citrain, Aur{\'e}lien and Corbett, Benjamin C and Corbett, James and others},
  journal={Journal of Open Source Software},
  volume={9},
  number={LLNL--JRNL-864747},
  year={2024},
  publisher={Lawrence Livermore National Laboratory (LLNL), Livermore, CA (United States)}
}

@article{li2025fiber,
  title={Fiber-optic seismic full waveform monitoring of groundwater dynamics},
  author={Li, Haipeng and Liu, Jingxiao and Mao, Shujuan and Yuan, Siyuan and Clapp, Robert G and Biondi, Biondo L},
  journal={Geophysical Research Letters},
  volume={52},
  number={20},
  pages={e2025GL117610},
  year={2025},
  publisher={Wiley Online Library}
}

@article{hu2023feasibility,
  title={Feasibility of seismic time-lapse monitoring of {CO$_2$} with rock physics parametrized full waveform inversion},
  author={Hu, Qi and Grana, Dario and Innanen, Kristopher A},
  journal={Geophysical Journal International},
  volume={233},
  number={1},
  pages={402--419},
  year={2023},
  publisher={Oxford University Press}
}

@article{mardan2023monitoring,
  title={Monitoring fluid saturation in reservoirs using time-lapse full-waveform inversion},
  author={Mardan, Amir and Giroux, Bernard and Fabien-Ouellet, Gabriel and Saberi, Mohammad Reza},
  journal={Geophysical Prospecting},
  volume={71},
  number={6},
  pages={1012--1029},
  year={2023},
  publisher={European Association of Geoscientists \& Engineers}
}

@article{di2025latent,
  title={Latent diffusion models for parameterization of facies-based geomodels and their use in data assimilation},
  author={Di Federico, Guido and Durlofsky, Louis J},
  journal={Computers \& Geosciences},
  volume={194},
  pages={105755},
  year={2025},
  publisher={Elsevier}
}

@article{bingham2019pyro,
  title={Pyro: Deep universal probabilistic programming},
  author={Bingham, Eli and Chen, Jonathan P and Jankowiak, Martin and Obermeyer, Fritz and Pradhan, Neeraj and Karaletsos, Theofanis and Singh, Rohit and Szerlip, Paul and Horsfall, Paul and Goodman, Noah D},
  journal={Journal of Machine Learning Research},
  volume={20},
  number={28},
  pages={1--6},
  year={2019}
}

@article{taufik2024learned,
  title={Learned regularizations for multi-parameter elastic full waveform inversion using diffusion models},
  author={Taufik, Mohammad H and Wang, Fu and Alkhalifah, Tariq},
  journal={Journal of Geophysical Research: Machine Learning and Computation},
  volume={1},
  number={1},
  pages={e2024JH000125},
  year={2024},
  publisher={Wiley Online Library}
}

@article{zhu2022integrating,
  title={Integrating deep neural networks with full-waveform inversion: Reparameterization, regularization, and uncertainty quantification},
  author={Zhu, Weiqiang and Xu, Kailai and Darve, Eric and Biondi, Biondo and Beroza, Gregory C},
  journal={Geophysics},
  volume={87},
  number={1},
  pages={R93--R109},
  year={2022},
  publisher={Society of Exploration Geophysicists}
}

@article{fang2020deep,
  title={Deep generator priors for {B}ayesian seismic inversion},
  author={Fang, Zhilong and Fang, Hongjian and Demanet, Laurent},
  journal={Advances in Geophysics},
  volume={61},
  pages={179--216},
  year={2020},
  publisher={Elsevier}
}

@article{li2020coupled,
  title={Coupled time-lapse full-waveform inversion for subsurface flow problems using intrusive automatic differentiation},
  author={Li, Dongzhuo and Xu, Kailai and Harris, Jerry M and Darve, Eric},
  journal={Water Resources Research},
  volume={56},
  number={8},
  pages={e2019WR027032},
  year={2020},
  publisher={Wiley Online Library}
}

@article{yin2024time,
  title={Time-lapse full-waveform permeability inversion: A feasibility study},
  author={Yin, Ziyi and Louboutin, Mathias and M{\o}yner, Olav and Herrmann, Felix J},
  journal={The Leading Edge},
  volume={43},
  number={8},
  pages={544--553},
  year={2024},
  publisher={Society of Exploration Geophysicists}
}

@article{jiang2024history,
  title={History matching for geological carbon storage using data-space inversion with spatio-temporal data parameterization},
  author={Jiang, Su and Durlofsky, Louis J},
  journal={International Journal of Greenhouse Gas Control},
  volume={134},
  pages={104124},
  year={2024},
  publisher={Elsevier}
}

@article{han2024surrogate,
  title={Surrogate model for geological {CO$_2$} storage and its use in hierarchical {MCMC} history matching},
  author={Han, Yifu and Hamon, Fran{\c{c}}ois P and Jiang, Su and Durlofsky, Louis J},
  journal={Advances in Water Resources},
  volume={187},
  pages={104678},
  year={2024},
  publisher={Elsevier}
}

@book{mavko2020rock,
  title={The rock physics handbook},
  author={Mavko, Gary and Mukerji, Tapan and Dvorkin, Jack},
  year={2020},
  publisher={Cambridge university press}
}

@inproceedings{xu2024diffsim,
  title={DiffSim: Denoising diffusion probabilistic models for generative facies geomodeling},
  author={Xu, Minghui and Song, Suihong and Mukerji, Tapan},
  booktitle={Fourth {I}nternational {M}eeting for {A}pplied {G}eoscience \& {E}nergy},
  pages={1660--1664},
  year={2024},
  organization={Society of Exploration Geophysicists and American Association of Petroleum}
}

@article{zunino2015monte,
  title={{Monte Carlo} reservoir analysis combining seismic reflection data and informed priors},
  author={Zunino, Andrea and Mosegaard, Klaus and Lange, Katrine and Melnikova, Yulia and Mejer Hansen, Thomas},
  journal={Geophysics},
  volume={80},
  number={1},
  pages={R31--R41},
  year={2015},
  publisher={Society of Exploration Geophysicists}
}

@article{laloy2017inversion,
  title={Inversion using a new low-dimensional representation of complex binary geological media based on a deep neural network},
  author={Laloy, Eric and H{\'e}rault, Romain and Lee, John and Jacques, Diederik and Linde, Niklas},
  journal={Advances in Water Resources},
  volume={110},
  pages={387--405},
  year={2017},
  publisher={Elsevier}
}

@article{lopez2021deep,
  title={Deep generative models in inversion: The impact of the generator's nonlinearity and development of a new approach based on a variational autoencoder},
  author={Lopez-Alvis, Jorge and Laloy, Eric and Nguyen, Frederic and Hermans, Thomas},
  journal={Computers \& Geosciences},
  volume={152},
  pages={104762},
  year={2021},
  publisher={Elsevier}
}

@article{wang2018conditioning,
  title={Conditioning {3D} object-based models to dense well data},
  author={Wang, Yimin C and Pyrcz, Michael J and Catuneanu, Octavian and Boisvert, Jeff B},
  journal={Computers \& Geosciences},
  volume={115},
  pages={1--11},
  year={2018},
  publisher={Elsevier}
}

@article{deutsch1992geostatistical,
  title={Geostatistical software library and user’s guide},
  author={Deutsch, Clayton V and Journel, Andre G and others},
  journal={New York},
  volume={119},
  number={147},
  pages={578},
  year={1992}
}

@article{manchuk2012flexible,
  title={A flexible sequential {Gaussian} simulation program: {USGSIM}},
  author={Manchuk, John G and Deutsch, Clayton V},
  journal={Computers \& Geosciences},
  volume={41},
  pages={208--216},
  year={2012},
  publisher={Elsevier}
}

@incollection{robert2004diagnosing,
  title={Diagnosing Convergence},
  author={Robert, Christian P and Casella, George},
  booktitle={{Monte Carlo} Statistical Methods},
  pages={363--413},
  year={2004},
  publisher={Springer}
}

@software{richardson_2023_8381177,
  author       = {Richardson, Alan},
  title        = {Deepwave},
  month        = sep,
  year         = 2023,
  publisher    = {Zenodo},
  version      = {v0.0.20},
  doi          = {10.5281/zenodo.8381177},
  url          = {https://doi.org/10.5281/zenodo.8381177},
}

@misc{li_2025_17917000,
  author       = {Li, Haipeng and Wang, Nanzhe and Durlofsky, Louis and Biondi, Biondo},
  title        = {Bayesian Full-waveform Monitoring of {CO$_2$} Storage with Fluid-flow Priors via Generative Modeling
                  },
  month        = dec,
  year         = 2025,
  publisher    = {Zenodo},
  version      = {1.0.0},
  doi          = {10.5281/zenodo.17917000},
  url          = {https://doi.org/10.5281/zenodo.17917000},
}

\end{document}